\documentclass{PoS}

\usepackage{float}
\RequirePackage{color}
\usepackage{colortbl}

\usepackage{multirow}
\usepackage{amsmath}
\usepackage{placeins}

\def\be{\begin{equation}} 
\def\ee{\end{equation}}
\def\bea{\begin{eqnarray}}
\def\eea{\end{eqnarray}}

\title{Dark Energy in String Theory\thanks{Includes work done in collaborations with Ed Hardy, Yessenia Olguin-Trejo, and Gianmassimo Tasinato.}}

\ShortTitle{Dark Energy in String Theory}

\author{Bruno Valeixo Bento\\
        Department of Mathematical Sciences, University of Liverpool, Liverpool L69 7ZL\\
        E-mail: \email{Bruno.Bento@liv.ac.uk}}

\author{Dibya Chakraborty\\
        Departamento de F\'isica, Universidad de Guanajuato, Loma del Bosque No. 103
        Col. Lomas del Campestre, C.P 37150 Le\'on, Guanajuato, M\'exico\\
        E-mail: \email{dibyac@fisica.ugto.mx}}
                                
\author{\speaker{Susha L. Parameswaran}\\
        Department of Mathematical Sciences, University of Liverpool, Liverpool L69 7ZL\\
        E-mail: \email{susha@liv.ac.uk}}

\author{Ivonne Zavala\\
        Physics Department, Swansea University, SA2 8PP, UK\\
        E-mail: \email{e.i.zavalacarrasco@swansea.ac.uk}}

\abstract{We consider various candidates for Dark Energy, motivated by string theory. Several no-go theorems push de Sitter string vacua, with $w=-1$, to the limits of theoretical control, and all known examples depend on a delicate interplay between different string theoretic ingredients.  On the other hand, runaway moduli directions are ubiquitous in string theory, and could plausibly source slow-roll quintessence.  We consider various candidate supergravity potentials, motivated by string theory,  including single-field K\"ahler potentials for bulk and local moduli, and leading superpotentials of the form $W = W_0 + A e^{-a \Phi}$ or $W = W_0 + A \Phi^p$.  Conditions on the scalar potential imposed by supergravity are very restrictive, ruling out e.g. quintessence with $K=-n\ln(\Phi+\bar{\Phi})$ and $W = W_0+A \Phi^p$.  Out of the examples considered, one can simultaneously satisfy $V>0$ and $\epsilon_V<1$ only for a deformation-like modulus with $K = k_0 + \frac{|\Phi|^{2n}}{k1}$ and a blow-up like modulus with $K=k_0 +\frac{(\Phi+\bar{\Phi})^{2n}}{k_1}$ when the leading order in the perturbative superpotential, $p$, is equal to $n$.  We also review the scenario of Thermal Dark Energy, where thermal effects in a light hidden sector hold a scalar field up away from the minimum of its zero-temperature potential.  This provides a viable model of Dark Energy with $w=-1$, consistent with known swampland conjectures, and motivates further early Thermal Dark Energy epochs with potentially observable consequences.}

\FullConference{Corfu Summer Institute 2019 "School and Workshops on Elementary Particle Physics and Gravity" (CORFU2019)\\
                31 August - 25 September 2019\\
                Corf\`u, Greece}

\begin{document}

\section{Introduction}
The nature of the Dark Energy that dominates our Universe today and drives its accelerated expansion remains an entirely open question, despite more than twenty years of effort from theorists since its discovery.  As we enter the next decade, a new generation of ground-based and satellite cosmological observations will probe the nature of Dark Energy to an unprecedented depth and precision.  Diverse and complementary techniques -- using supernovae, baryon acoustic oscillations, galaxy cluster abundances and weak gravitational lensing -- will map the expansion history of the Universe and its influence on structure formation, and thus constrain the equation of state, time-dependence and couplings that characterise Dark Energy.

So far, observations are consistent with a Dark Energy equation of state $w=-1$, that is, a cosmological constant.  However, the dark energy density is around 120 orders of magnitude smaller than naive estimates of the cosmological constant from quantum field theory (for a review see \cite{Martin:2012bt}).  Especially because of its connection to the cosmological constant problem, one might expect that the fundamental nature of Dark Energy has to be addressed within an ultraviolet complete theory of quantum gravity.  Within string theory, Dark Energy cannot be understood without understanding supersymmetry breaking and moduli stabilisation. It has long been known that moduli stabilisation to a de Sitter vacuum would be challenging.  Dine and Seiberg \cite{Dine:1985he} had noted already in the early days of string theory that, in the perturbative regimes of the string coupling and $\alpha'$ expansions, the dilaton and volume moduli must be runaway directions, unless there are parameters available to tune; only by fine-tuning terms at different orders in the perturbative expansions against each other can a local minimum be obtained.  Moreover, Gibbons \cite{Gibbons:1984kp} and Maldacena and N\'u\~nez \cite{Maldacena:2000mw} proved that two-derivative supergravity with only positive tension objects does not admit de Sitter solutions.  There have been many subsequent extensions to this no-go theorem, including for example the CFT computation that rules out de Sitter vacua in heterotic string theory to all orders in $\alpha'$ (but at string tree level) \cite{Kutasov:2015eba}.   

Of course, an essential ingredient of any no-go theorem is its starting assumptions, and the no-go theorems just mentioned may well provide clues as to how the obstructions might be evaded.  The last two decades have seen considerable advances in this direction, beginning with the seminal papers on flux compactifications \cite{Giddings:2001yu} and KKLT \cite{Kachru:2003aw}.  The large number of possible string vacua expected led to the picture of the String Theory Landscape \cite{Susskind:2003kw}, which ought to include our own Universe with a tiny positive cosmological constant.  Put together with a means to populate the Landscape, such as eternal inflation, this could give rise to an anthropic explanation for a finely-tuned cosmological constant, an idea that led quite remarkably to an early prediction of Dark Energy in \cite{Weinberg:1988cp}.

Several scenarios of de Sitter string theory vacua have thus been proposed, in all the different duality regimes of string theory (for recent reviews from different perspectives see \cite{Danielsson:2018ztv} and \cite{Cicoli:2018kdo}).  Much technical progress has been made towards making explicit realisations of these scenarios.  However, it is fair to say that these constructions are always at -- or beyond -- the limits of theoretical control.  These difficulties have led to the conjecture that metastable de Sitter vacua may actually be impossible to embed consistently in an ultraviolet complete theory of quantum gravity \cite{Obied:2018sgi, Garg:2018reu, Ooguri:2018wrx}.  In other words, effective field theories with metastable de Sitter vacua may belong to the String Theory Swampland, and rather than googol-like numbers of de Sitter vacua there may be none at all. The de Sitter Swampland Conjecture has intriguing connections to other swampland conjectures \cite{Ooguri:2018wrx, Bedroya:2019snp} and wider discussions around the quantum aspects of de Sitter \cite{Witten:2001kn, Banks:2012hx, Maltz:2016iaw, Dvali:2018jhn}.

If string theory does not have metastable de Sitter vacua, then there must be some other explanation for Dark Energy.  And even if metastable de Sitter vacua are allowed within string theory, it is important to consider other well-motivated alternatives, and how they might be observationally distinguishable.  There has been surprisingly little work towards alternative Dark Energy models within string theory (see \cite{Cicoli:2018kdo} for a review).  In quintessence models, the accelerated expansion is driven by an extremely light, slowly rolling field (see \cite{Tsujikawa:2013fta} for a review).  These models not only have the challenge of fine-tuning (see \cite{Hebecker:2019csg} for a recent discussion in the context of string theory and supergravity), but must also somehow evade bounds from fifth forces and time-variation of fundamental constants.  Amongst the most attractive quintessence candidates are therefore string axions, as current fifth forces experiments do not probe the spin-dependent couplings to which pseudoscalars give rise.  However, slow-roll axion quintessence requires scenarios like super-Planckian axion decay constant \cite{Freese:1990rb,Svrcek:2006hf} (difficult to find in string theory \cite{Banks:2003sx} and in tension with swampland conjectures \cite{ArkaniHamed:2006dz}), a delicate interplay of ingredients to realise alignment \cite{Kim:2004rp}, axion monodromy \cite{McAllister:2014mpa} or periodic plateaus \cite{Parameswaran:2016qqq}, or a fine-tuning of initial conditions at hilltops \cite{Cicoli:2018kdo}.  Thus string axion quintessence is no more under control than string de Sitter vacua.  Other quintessence candidates that have been considered are closed-string moduli associated with internal four-cycles \cite{Cicoli:2012tz}, but again these are based on intricate moduli stabilisation scenarios similar to those for de Sitter vacua.

On the other hand, given the ubiquity of runaway moduli in string compactifications, one may expect runaway quintessence scenarios to be very naturally realised, without the need for delicate interplay between various not-so-under-control ingredients.  Moreover, if the runaway direction were a local modulus in a hidden sector, it may plausibly avoid constraints from fifth forces and time variation of fundamental constants.  However, as we will discuss, it is intriguingly challenging to identify runaway directions in string theory that are sufficiently flat to source acceleration.  The simplest Dine-Seiberg runaway -- a supersymmetric flat direction lifted by non-perturbative effects -- turns out to be always too steep \cite{Olguin-Tejo:2018pfq}.  We also consider several simple, well-motivated perturbative runaways within four-dimensional supergravity, and find that constraints from supergravity are very restrictive, almost always rendering the runaways either too steep or with negative potential energy.  Interestingly, there is a window in parameter space for K\"ahler potentials resembling those for local string moduli, $K = k_0 + \frac{(\Phi+\bar{\Phi})^{2n}}{k_1}$ or $K = k_0 + \frac{|\Phi|^{2n}}{k_1}$, but only when the order of the leading superpotential coupling, $W=W_0 + A \Phi^p$, is restricted to $p=n$.  It would of course be extremely interesting to find concrete string theory constructions that satisfy these conditions.

The lack of any compelling model of de Sitter or quintessence from string theory thus far, drives the search for completely novel string theory origins for Dark Energy.  In Thermal Dark Energy \cite{Hardy:2019apu}, high temperature effects in a light hidden sector hold a string theory modulus away from its zero temperature minimum.  This sources a positive potential energy density that mimics a cosmological constant, until, as the temperature falls, a phase transition to the true minimum takes place.  Thermal Dark Energy is consistent with all swampland conjectures, and has distinctive experimental and observational signatures, including a possible resolution to the $H_0$-tension.

In this talk, we will first briefly review the status of de Sitter vacua in string theory.  The aim of this section is not to do justice to the vast body of work on this topic, but to illustrate the technical challenges involved, and show that -- with present knowledge -- we have to be agnostic on the existence or not of de Sitter vacua in string theory.   We will then discuss quintessence models in string theory, focusing on the runaway potentials that are ubiquitous in string compactifications, summarising \cite{Olguin-Tejo:2018pfq} and including some further results using supergravity.  Finally we will review Thermal Dark Energy \cite{Hardy:2019apu}.

\section{De Sitter vacua in string theory}
The search for de Sitter string compactifications is typically made via four-dimensional effective field theories, by looking for a scalar potential for the light moduli that has a positive-definite metastable minimum, $\langle V(\phi_i) \rangle_{min} >0$.  It is important to note that finding such a de Sitter vacuum does not alone address the cosmological constant problem, and the total vacuum energy would typically be of the form $\Lambda \approx \langle V(\phi_i) \rangle_{min} + M_{kk}^4$.  Put another way, any classical or quantum de Sitter vacuum that is found within a four-dimensional effective field theory, would generally be modified by $g_s$, $\alpha'$ and heavy-state corrections, which contribute some $\Delta V$ to the vacuum energy.  With sufficiently large numbers of different $\langle V(\phi_i) \rangle_{min}$ and $\Delta V$, a fine-tuned cancellation can allow $\Lambda \sim 10^{-121}M_{pl}^4$.  With this caveat in place, we now briefly review the status of de Sitter string vacua.  

Type IIA compactifications, having both even and odd-form fluxes, have the potential to stabilise all moduli classically, and the Gibbons-Maldacena-Nu\~nez no-go theorem could be evaded due to the presence of negative tension orientifold planes, essential for tadpole cancellation.  However, it was shown early on that for type IIA Calabi-Yau orientifolds with standard fluxes, $\frac{|\nabla V|}{V} \geq \sqrt{\frac{54}{13}}$, thus ruling out de Sitter vacua in this class of compactifications \cite{Hertzberg:2007wc} (see \cite{Wrase:2010ew} and \cite{Obied:2018sgi} for generalisations).  Type IIA compactifications on negative curvature manifolds are more promising.  De Sitter solutions have been found using a non-zero Romans mass, negatively curved spaces and O6 planes \cite{Caviezel:2008tf, Flauger:2008ad}.  However -- aside from control issues e.g. from smearing sources and using unquantized fluxes -- all de Sitter solutions found have turned out to have tachyonic instabilities \cite{Roupec:2018mbn}.

On another point of the duality star, heterotic orbifold compactifications lend a large amount of control thanks to the fact that the worldsheet theory is free, so much of the low energy effective field theory can be directly computed up to a few free parameters (see \cite{Parameswaran:2010ec} for a review).  Again, although both orbifold and smooth scenarios have been proposed to realise de Sitter vacua by including non-perturbative efffects \cite{Parameswaran:2010ec, Anderson:2011cza, Cicoli:2013rwa}, those explicit top-down de Sitter vacua \cite{Parameswaran:2010ec} that have been found are all tachyonic.  As for the explicit type IIA compactifications \cite{Danielsson:2012et}, one technical challenge is the large number of light moduli, with ten real degrees of freedom.  On one hand, a large number of fields can help to arrive at a de Sitter solution, as each supersymmetry breaking contribution gives a positive definite contribution to the potential energy.  However, with a random scan for stationary points in the ten-field scalar potential, there is only a one in $2^{10}$ chance of finding one with not a single negative eigenvalue in the Hessian matrix.  Moreover, concrete examples considered in \cite{Parameswaran:2010ec} have only 13 free parameters available to tune for the 10 equations and 11 inequalities that give conditions for a metastable de Sitter minimum.  Another interesting feature of heterotic orbifold compactifications is how target-space modular symmetries acting on the geometric moduli constrain the scalar potential, such that the potential depends on the moduli as certain modular forms.  Studies of general modular invariant potentials have identified de Sitter saddle points, but no de Sitter minima \cite{Olguin-Tejo:2018pfq, Gonzalo:2018guu}.

Arguably the most studied scenarios for de Sitter vacua in string theory are in type IIB.  There, fluxes can stabilise the dilaton and complex structure, but some interplay between classical, perturbative and non-perturbative effects -- in $g_s$ and $\alpha'$ -- are necessary to stabilise the K\"ahler moduli.  KKLT \cite{Kachru:2003aw} evades the Dine-Seiberg runaway by fine-tuning classical effects against non-perturbative effects, whereas the Large Volume Scenario \cite{Balasubramanian:2005zx} exploits the fact that there are two perturbative expansions, so the leading terms in each perturbative expansion can balance against each other.  In both cases, these interplays lead to an anti-de Sitter vacuum, which is uplifted to de Sitter with further ingredients like probe anti-D3-branes \cite{Kachru:2003aw}, K\"ahler corrections \cite{Westphal:2006tn}, or T-branes \cite{Cicoli:2015ylx}.  Further, examples of directly obtaining de Sitter vacua without any uplifting based on perturbative and non-perturbative contributions were presented in \cite{Blaback:2013qza}. 

Out of these proposed ingredients for de Sitter, anti-D3-branes are the most well understood, but still raise many questions, such as how their backreaction affects the background solution and the stabilisation of geometric moduli (e.g \cite{Bena:2018fqc}).  Another question is how intrinsically four-dimensional non-perturbative effects can themselves be used to stabilise an otherwise rolling four-dimensional compactification \cite{Sethi:2017phn}.  
KKLT and the Large Volume scenarios have undergone a large amount of scrutiny in recent years, and whether or not the proposals can be consistently realised within string theory is still under debate.

Other interesting but not-yet-sufficiently-developed proposals for string theory de Sitter vacua include non-geometric fluxes \cite{Blaback:2013ht}. For all scenarios, the holy grail would be a rigorous ten-dimensional solution with four-dimensional de Sitter space and compact internal dimensions (see e.g. \cite{Kachru:2019dvo, Gautason:2019jwq,Grana:2020hyu}  and \cite{Cordova:2018dbb,Cordova:2019dbb} for recent work towards this objective, in type IIB and IIA, respectively), together with a stability analysis.  These solutions may need to include some time-dependence (see e.g. \cite{Dasgupta:2019rwt}).

\section{Runaway quintessence}
Given the challenges in constructing metastable de Sitter solutions in string theory, and the genericity of Dine-Seiberg runaway directions in moduli space, it is very natural to suppose that Dark Energy might be due to a slowly rolling or frozen runaway quintessence field.  The immediate challenges are (i) the fine-tuning of the vacuum energy is now extended to the fine-tuning of the very light mass of the quintessence field, $m \lesssim 10^{-33}$eV (ii) such a light scalar field would usually mediate unobserved fifth forces, in particular any scalar with mass less than around meV must have weaker than Planckian couplings to the Standard Model fields (iii) as scalar fields in string theory usually correspond to couplings, the rolling quintessence field can lead to time-variation of fundamental constants.  Problems (ii) and (iii) may be addressed if the quintessence field comes from a hidden sector that is sequestered from the visible sector, although there are as yet no explicit constructions that realise the degree of sequestering that would be necessary (see \cite{Berg:2010ha} for some challenges, \cite{Aparicio:2014wxa} for an more optimistic point of view, and \cite{Acharya:2018deu} for recent work in this direction).

Arguably the simplest class of runaway moduli are those originating from supersymmetric flat directions.  We will now show, however, that it is impossible for such a runaway tail to play the role of Dark Energy and source an accelerated expansion.  Then we consider other possible runaway moduli.

Suppose some early Universe scenario, such as inflation, which ends in a supersymmetric Minkowski minimum in which most of the string moduli, $\Phi_i$, are stabilised and heavy:
\be
\langle D_i W_{susy} \rangle = 0 \quad \textrm{and} \quad \langle W_{susy} \rangle = 0 \,.
\ee
Assume for simplicity a single flat direction, corresponding to the chiral superfield, $\Phi$, with scalar component also labelled $\Phi = \phi + i \theta$, with saxion $\phi$ and axion $\theta$. For example, a bulk modulus may have leading K\"ahler potential:
\be
K = -n\ln(\Phi + \bar{\Phi}) \,, \label{E:K}
\ee
with e.g. $n=3$ for the overall volume modulus, or $n=1$ for another K\"ahler modulus, a complex structure or dilaton, or some other $n<3$ for a fibre modulus.

As we are in a supersymmetric Minkowski background, the flat direction is protected to all finite orders by the non-renormalisation theorem: a Peccei-Quinn shift symmetry forbids dependence on $\theta$, and holomorphicity of the superpotential implies dependence on $\phi$ is also forbidden \cite{Dine:1986vd} (see \cite{Burgess:2005jx} and \cite{GarciadelMoral:2017vnz} for interesting generalisations). Even if $K$ can receive perturbative corrections, the flat direction cannot be lifted so long as $W=0$.  The Peccei-Quinn shift symmetry is broken by non-perturbative effects, so at some scale, say before BBN, a leading order non-perturbative superpotential is generated:
\be
W = A e^{-a \Phi} \,.
\ee
which leads to a scalar potential:
\be
V = \frac{A^2}{2^n n ~M_{pl}^2} e^{-2 a\phi} \phi^{-n}\left(n^2 + 4 a^2 \phi^2+n\left(-3+4\,a\,\phi\right)\right) \,. \label{E:pot}
\ee

\begin{figure}[t!]
\begin{center}
\includegraphics[width=0.45\textwidth]{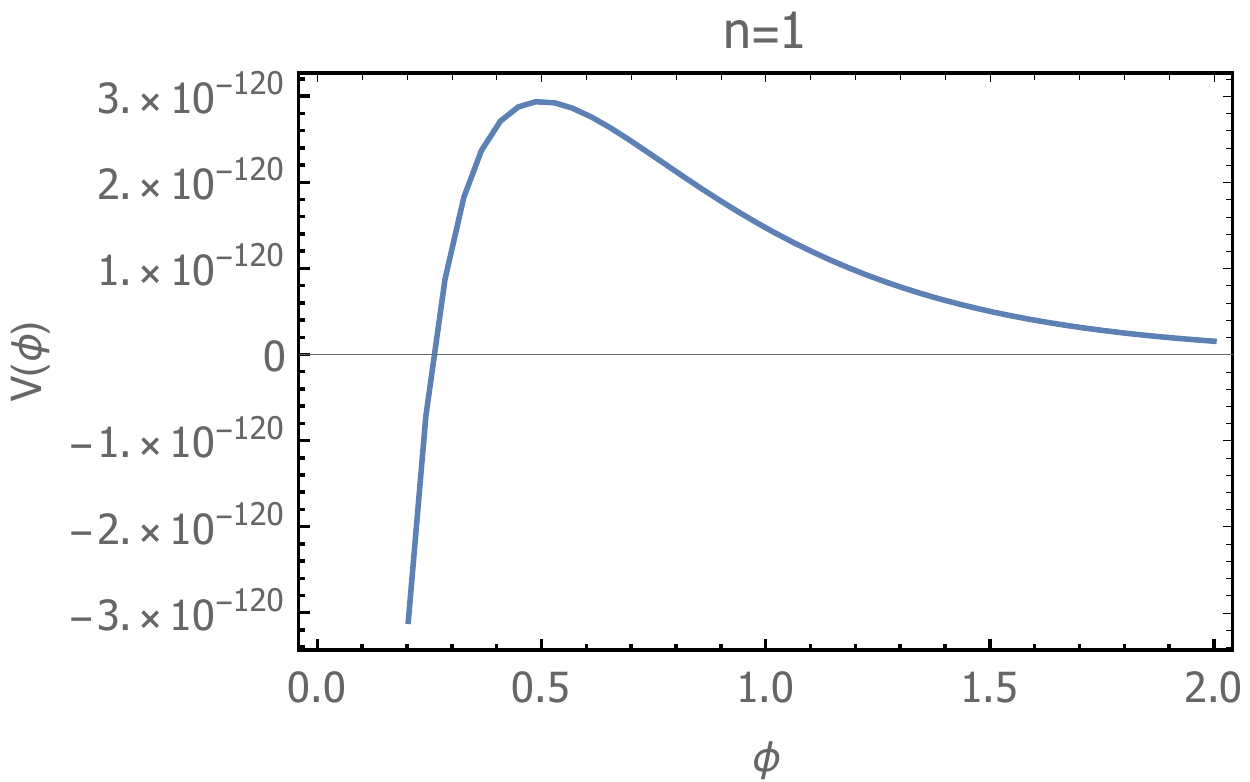} \hspace{1cm} \includegraphics[width=0.45\textwidth]{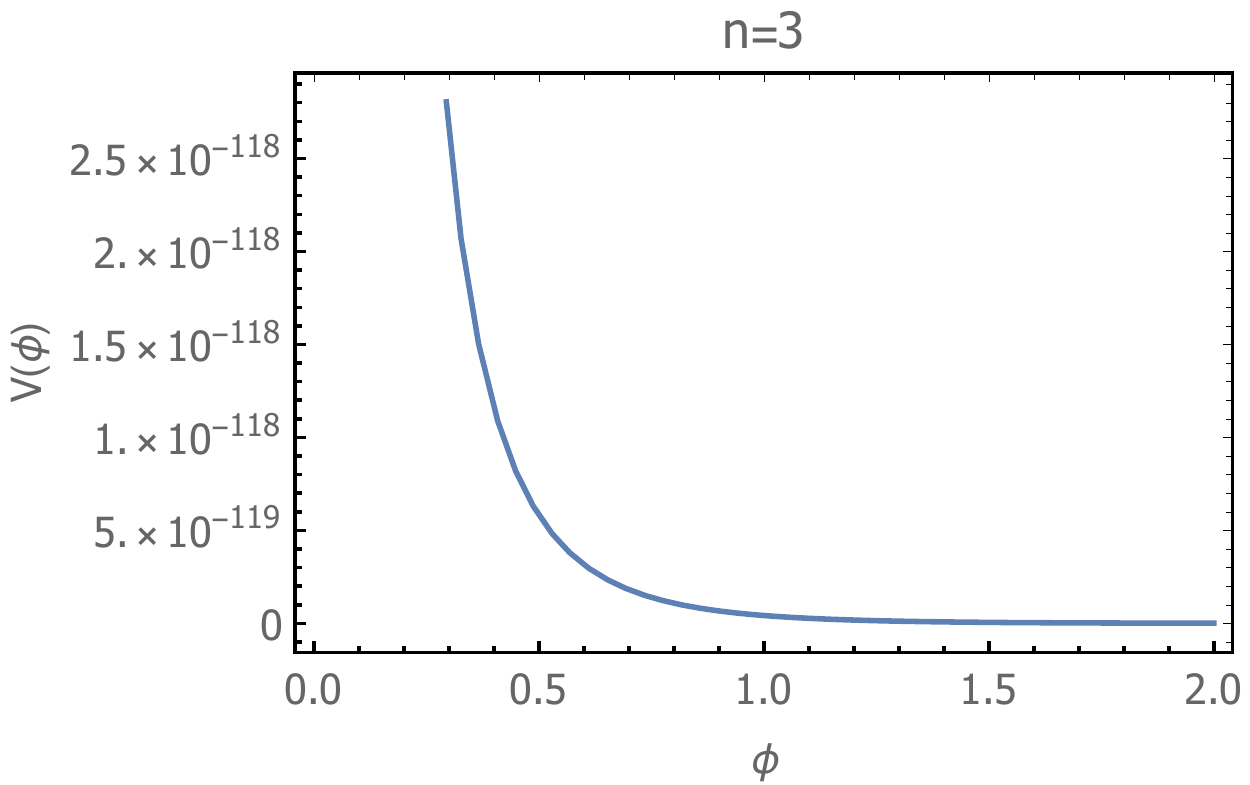}
\end{center}
\caption{Potential (\ref{E:pot}) for $n=1$ (left) and $n=3$ (right) with $a=\sqrt{2}$ and $A = e^{-1105/8}$ in Planck units.\label{F:pot}}
\end{figure} 
Thus the flat direction for $\phi$ is lifted.  The axion, $\theta$, at this leading order remains a flat direction, but will be lifted by subleading corrections.
The overall scale of the potential energy is fixed by the constant $A$, and the exponential suppression in $\phi$.  In a complete string theory model, $A$ would depend on the heavy moduli, $\Phi_i$, and it could itself be exponentially suppressed in those heavy moduli.  For example, in gaugino condensation in a hidden sector, $W = \mu^2 e^{-a f}$, where $\mu$ is the scale at which gaugini condense, $a$ is determined from the hidden sector beta function coefficient, and the gauge kinetic function is given by $f = \Phi + \Delta_{1-loop}(\Phi_i)$, with one-loop threshold corrections depending on heavy moduli.   So it should not be difficult to obtain a non-perturbatively generated potential energy density of order $\Lambda \sim e^{-280} M_{pl}^4$.

However, it is very simple to show that such a scalar potential cannot source an accelerated expansion at its tail.  Indeed, as is well known, the Friedman and Klein-Gordon equations conspire together to reveal that for a frozen or slowly rolling field to dominate the Universe (over matter and radiation) and drive an accelerated expansion, the slow-roll condition:
\be
\epsilon_V = \frac{M_{pl}^2}{2} g^{\phi\phi}\left(\frac{V'(\phi)}{V(\phi)}\right)^2 < 1  \label{E:sr}
\ee
must be satisfied.  Note that so long as the field is frozen by Hubble friction, it mimics effectively a cosmological constant, but as the Hubble parameter falls the field will eventually begin to roll.
Plugging (\ref{E:pot}) into the slow-roll condition (\ref{E:sr}), and considering the tail of the runaway, $\phi\rightarrow \infty$, one finds:
\be 
\epsilon_V \rightarrow  \frac{4}{n} a^2 \phi^2 \quad \textrm{as} \quad \phi \rightarrow \infty \,.
\ee
Therefore it is impossible to satisfy the slow-roll condition $\epsilon_V <1$ (\ref{E:sr}) and drive an accelerated expansion at the tail of the non-perturbative runaway. Another way to state this is that the scalar potential for the canonically normalised field, $\varphi = M_{pl}\sqrt{\frac{n}{2}}\log\phi$:
\be
V(\varphi) \approx \frac{4 A^2 a^2}{2^n n ~M_{pl}^2} e^{-2a e^{\sqrt{\frac{2}{n}}\frac{\varphi}{M_{pl}}}} \left(e^{\sqrt{\frac{2}{n}}\frac{\varphi}{M_{pl}}}\right)^{2-n} \quad \textrm{at large }\varphi,
\ee
with its double-exponential dependence, is too steep to allow a slow-roll accelerated expansion.  Note that with suitably fine-tuned initial conditions, it is possible to obtain viable models of frozen quintessence at the hilltop of the potential $V(\phi)$ with $n=1$, seen in Figure \ref{F:pot} (a fine-tuning of around 4\% for the parameter space studied in \cite{Olguin-Tejo:2018pfq} is sufficient).  However, it is difficult to find an explanation for such special initial conditions, even an anthropic one.

Perturbative runaway directions may seem more promising.  However, consider for example the K\"ahler potential (\ref{E:K}) with a perturbative superpotential (later we will extend the superpotential to $W = W_0 + A \Phi^p$):
\be
K = -n\ln(\Phi+\bar{\Phi}) \quad \textrm{and} \quad W = A \Phi^p\,,
\ee
for which:
\be
V(\phi) = \frac{A^2}{2^n n ~M_{pl}^2} \phi^{-n}(\phi^2+\theta^2)^{-1+p}\left(\left(\left(-3+n\right)n-4np+4p^2\right)\phi^2 + (-3+n)n\theta^2 \right) \,.
\ee
It is useful to consider the potential around $\theta\approx 0$:
\bea
V(\phi) &=& \frac{A^2}{2^n n ~M_{pl}^2}  \left((-3+n)n-4np+4p^2\right)\phi^{-n+2p}\nonumber \\ && + \frac{A^2}{2^n n ~M_{pl}^2}  p \left((-3+n)n-4np+4p^2 +4(n-p)\right)\phi^{-2-n+2p} \theta^2 + \mathcal{O}(\theta^3)
\eea
It is easy to see that $\theta=0$ corresponds to a metastable minimum as long as 
\be
p < \frac{n+1}{2} - \frac{\sqrt{n+1}}{2}
\quad \textrm{or} \quad
p > \frac{n+1}{2} + \frac{\sqrt{n+1}}{2}
\ee
while the scalar potential $V(\phi)>0$ at $\theta=0$ provided
\be 
p < \frac{n}{2} - \frac{\sqrt{3n}}{2} 
\quad \textrm{or} \quad
p > \frac{n}{2} + \frac{\sqrt{3n}}{2} 
\ee 
For example, if $n=1$ and $p\geq 2$, then $\theta$ is stabilised at $\theta=0$, and the potential for the canonically normalised $\varphi$ becomes $V(\varphi) = \frac{A^2}{2 M_{pl}^2}  (-2-4p+4p^2)e^{(-1+2p)\sqrt{2}\varphi/M_{pl}}>0$ with, however, a slow-roll parameter $\epsilon_V = (1-2p)^2$ always greater than one.
For general $n>0$ and $p$, assuming $\theta=0$, the slow-roll parameter is:
\be
\epsilon_V = \frac{\left(n-2p\right)^2}{n} \,, 
\ee
and it is straightforward to show that it is impossible to have simultaneously $\epsilon_V < 1$ and $V(\phi) > 0$.

If we relax somewhat the constraints from a supergravity description of the action, a K\"ahler potential (\ref{E:K}) implies that any power-law scalar potential for $\phi$, $V(\phi) = A \phi^{-p}$ will lead to a standard exponential scalar potential for the canonically normalised field:
\be
V(\varphi) = A e^{-\sqrt{\frac{2p^2}{n}} \frac{\varphi}{M_{pl}}} \,. \label{E:expquint}
\ee
We've seen above that supergravity imposes relations within the perturbative scalar potential, between its linear-coefficient and the coefficient in the exponent, such that it is impossible to have both slow-roll $\epsilon_V<1$ and $V(\varphi) > 0$.  However, for (\ref{E:expquint}), a slow-roll accelerated expansion is possible for:\footnote{See \cite{Copeland:1997et,Copeland:2006wr} for a dynamical systems analysis of such potentials.  In \cite{Agrawal:2018own}, observational constraints on the dark energy equation of state $w(z)$ were used to constrain the constant in the de Sitter swampland constraint $|\nabla V| \gtrsim c V$ to $c \lesssim 0.6$.  Here, we consider the simplest scenario of a frozen field mimicking a cosmological constant, $w=-1$, for most of the cosmological history.}
\be
\frac{p^2}{n} \lesssim 1 \,.
\ee
For the value $n=1$ often seen for bulk string moduli, then we would need $p \lesssim 1$.  A fibre modulus with e.g. $n=2$ \cite{Cicoli:2009fibre} would need $p \lesssim \sqrt{2}$.  It would be very interesting to identify such perturbative runaways in explicit, well-under-control string constructions, though the lack of supersymmetry might make this particularly difficult.  If successful, one would then have to furthermore explain how the hierarchy in the vacuum energy and mass are stable with respect to the ultraviolet cutoff, and how to avoid fifth forces (note that, even if not sequestered, fundamental constants would not vary with time so long as the quintessence field is frozen by Hubble friction).

\begin{figure}[t!]
\begin{center}
\includegraphics[width=0.6\textwidth]{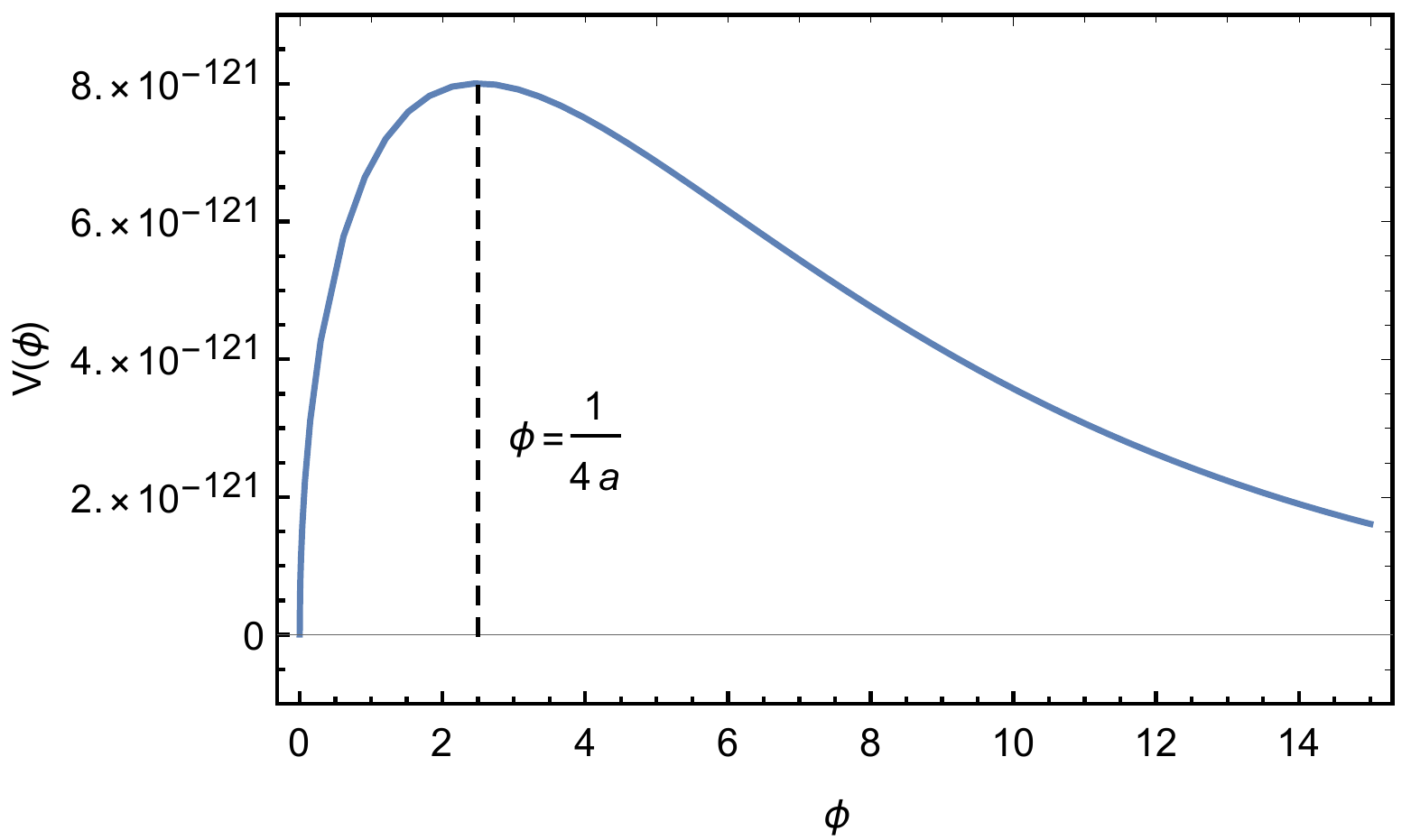}
\end{center}
\caption{Potential arising from blow up modulus with $K$ given in (\ref{E:blowupK}) and a non-perturbative superpotential, $W = A e^{-a\Phi}$, for $k_0 = -265$, $k_1 = 2075$, $k_2 = 1$, $A=1.5$ and $a = 0.1$, in Planck units.\label{F:potblowup}}
\end{figure}

Bearing in mind the need to suppress fifth forces, it is interesting to consider a local modulus, which may be sequestered from the Standard Model using geometric separation within the extra dimensions.  However, we will now see that the simplest models within supergravity again do not allow for slow-roll quintessence.  Consider e.g. a blow-up modulus with K\"ahler potential (see e.g. \cite{Cicoli:2008va} for explicit string examples of such moduli):
\be
K = k_0-2\ln(k_1 - k_2 (\Phi + \bar{\Phi})^{3/2}) \approx k_0 - 2 \ln(k_1) + 2 \frac{k_2}{k_1} (\Phi+\bar{\Phi})^\frac32, \label{E:blowupK}
\ee
where in the $\approx$ we assumed small values of the blow-up modulus, $\frac{k_2}{k_1} (\Phi + \bar{\Phi})^{3/2} \ll 1$.  Then, the canonically normalised field for the blow-up modulus is:
\be
\varphi = \frac{2^{7/4}}{\sqrt{3}}\sqrt{\frac{k_2}{k_1}} \phi^\frac{3}{4} \,.
\ee
Consider a non-perturbative superpotential, $W = A e^{-a\Phi}$.  The full scalar potential is:
\be 
V(\phi) = \frac{A^2}{3k_1^2 M_{pl}^2} e^{k_0 - 2a\phi + 2\frac{k_2}{k_1}(2\phi)^{3/2}} \left(-9 + 8(a\phi) \frac{a\phi}{\frac{k_2}{k_1}(2\phi)^{3/2}} - 24(a\phi) + 18\frac{k_2}{k_1}(2\phi)^{3/2}\right)
\ee 
When $a \phi < \frac{k_2}{k_1} (2\phi)^\frac32 \ll 1$ or $\frac{k_2}{k_1} (2\phi)^\frac32 < a\phi \ll 1$ this leads to negative potential energy:
\be
V(\phi) \approx -\frac{3 A^2 e^{k_0}}{k_1^2 ~M_{pl}^2} \,.
\ee
When instead $a \phi \gtrsim 1$, the potential has an exponential dependence in $\phi$:
\be
V(\phi) \approx \frac{8 A^2}{3 k_1^2 M_{pl}^2} e^{k_0 - 2 a \phi} \frac{k_1}{k_2 (2\phi)^{3/2}} a^2 \phi^2 \,.
\ee
The slow-roll parameter $\epsilon_V$ for the latter case is:
\be
\epsilon_V = \frac{k_1}{2 k_2 (2 \phi)^{3/2}}\frac23(1-4a\phi)^2,
\ee
so $\epsilon_V > 1$, and slow-roll quintessence is impossible, unless $\phi \sim \frac{1}{4a}$, which corresponds to fine-tuning the initial value of $\phi$ to the hilltop, shown in Figure \ref{F:potblowup}.

Another candidate amenable to sequestering would be a local modulus corresponding to the deformation parameter at the tip of a conifold, which has K\"ahler potential of the form \cite{Douglas:2007tu}:
\be
K = k_0 + k_1 |\Phi|^2\left(\ln\left(\frac{k_2}{|\Phi|}\right)+1\right)+k_3 |\Phi|^{\frac23}\,, 
\ee
and a flux-generated superpotential:
\be
W= -i w_1 \Phi \left(\ln \left(\frac{k_2}{\Phi}\right)+1 \right) + i w_2 \Phi \,.
\ee
For example, assuming a highly-warped scenario with $|\Phi| \ll l_s^3$, the $k_3$ term dominates over the $k_1$ term in $K$.  The slow-roll parameter turns out to be:
\be
\epsilon_V \approx \frac{16}{3}\left(1 + \frac{3}{4k_3 |\Phi|^{\frac23}}\right) > 1 \,.
\ee
More generally, a local modulus with K\"ahler potential of the form $K = k_0 + \frac{|\Phi|^{2 n}}{k_1}$ or $K = k_0 + \frac{(\Phi + \bar{\Phi})^{2n}}{k_1}$ with a non-perturbative superpotential $W = A e^{-a \Phi}$ (again, below we will extend this to $W = W_0+A e^{-a \Phi}$) leads, respectively, to the scalar potentials with exponential envelopes:
\bea
V(\phi) &=& \frac{A^2}{n^2 ~M_{pl}^2} e^{k_0 - 2 a \phi + \frac{\phi^{2n}}{k_1}}
\left((a\phi)^2 \frac{k_1}{\phi^{2n}} 
- 3n^2 - 2n(a\phi)
+ n^2 \frac{\phi^{2n}}{k_1} \right) \quad \textrm{for }\theta=0 \nonumber \\
&\approx&  \frac{A^2}{n^2 ~M_{pl}^2}  e^{k_0 - 2 a \phi}\left(a^2 \phi^2\right)\frac{k_1}{\phi^{2n}} \quad \textrm{when }\phi^{2n} \ll k_1\textrm{ and }a \phi \gtrsim \frac{\phi^{2n}}{k_1}
\eea
or
\bea
V(\phi) &=& \frac{A^2}{(2n-1)2n ~M_{pl}^2}  e^{k_0-2a\phi+ \frac{(2\phi)^{2n}}{k_1}} 
\left(4(a\phi)^2 \frac{k_1}{(2\phi)^{2n}} 
- 6n(2n-1) - 8n(a\phi) 
+ 4n^2 \frac{(2\phi)^{2n}}{k_1}\right) \nonumber\\
&\approx& \frac{A^2}{(2n-1)2n ~M_{pl}^2}  e^{k_0-2a\phi} \left(4a^2 \phi^2\right)\frac{k_1}{(2\phi)^{2n}} \quad \textrm{when }(2\phi)^{2n} \ll k_1\textrm{ and }a \phi \gtrsim \frac{(2\phi)^{2n}}{k_1} \,,
\eea
(for $a \phi < \frac{\phi^n}{k_1} \ll 1$, we instead have a negative potential energy $V(\phi) \approx - \frac{3 A^2 e^{k_0}}{M_{pl}^2}$ for both potentials).  The corresponding approximate slow-roll parameters are, respectively:
\be
\epsilon_V \approx \frac{(n-1+a\phi)^2}{n^2} \frac{k_1}{\phi^{2n}} > 1
\ee
or
\be
\epsilon_V \approx \frac{2(n - 1 + 2 a \phi)^2}{(2n-1)n}\frac{k_1}{(2\phi)^{2n}} > 1 \,.
\ee
Instead, a local modulus with K\"ahler potential of the form $K = k_0 + \frac{|\Phi|^{2 n}}{k_1}$ or $K = k_0 + \frac{(\Phi + \bar{\Phi})^{2n}}{k_1}$ with a perturbative superpotential $W = A \Phi^p$ (extended to $W = W_0+ A \Phi^p$ below) leads, respectively, to the power-law scalar potentials:
\bea
V(\phi) &=& \frac{A^2}{n^2 ~M_{pl}^2}e^{k_0 + \frac{\phi^{2n}}{k_1}}
\phi^{2p}\frac{k_1}{\phi^{2n}}\left(p^2 - n(3n - 2p)\frac{\phi^{2n}}{k_1} + n^2\frac{\phi^{4n}}{k_1^2}\right)\quad \textrm{for }\theta=0 \nonumber \\
&\approx&  \frac{A^2e^{k_0}}{n^2 ~M_{pl}^2}\frac{k_1}{\phi^{2n}}\phi^{2p}p^2 \quad \textrm{when }\phi^{2n} \ll k_1
\eea
or
\bea
V(\phi) &=& \frac{A^2}{(2n-1)2n ~M_{pl}^2} e^{k_0 + \frac{(2\phi)^{2n}}{k_1}}\phi^{2p}\frac{k_1}{(2\phi)^{2n}}
\left(4p^2 + 2n(3-6n+4p)\frac{(2\phi)^{2n}}{k_1} + 4n^2\frac{(2\phi)^{4n}}{k_1^2}\right) \nonumber \\
&\approx&  \frac{A^2e^{k_0}}{(2n-1)2n ~M_{pl}^2} \frac{k_1}{(2\phi)^{2n}}\phi^{2p}(4p^2) \quad \textrm{when }(2\phi)^{2n} \ll k_1 \,.
\eea
In the first case, the deformation-like modulus, the slow-roll parameter is:
\bea
\epsilon_V &=& \frac{k_1}{\phi^{2n}} \frac{\left(p^2(p-n) + 3n(p-n)\frac{\phi^{2n}}{k_1} - n^2(2n-3p)\frac{\phi^{4n}}{k_1^2} + n^3\frac{\phi^{6n}}{k_1^3}\right)^2}{n^2\left(p^2 - n(3n-2p)\frac{\phi^{2n}}{k_1} + n^2\frac{\phi^{4n}}{k_1^2}\right)^2}\\ \nonumber
&=&\frac{(n-p)^2}{n^2}\frac{k_1}{\phi^{2n}} + \mathcal{O}\left(\frac{\phi^{2n}}{k_1}\right) 
\quad \textrm{or} \quad \frac{\phi^{6n}}{k_1^3} + \mathcal{O}\left(\frac{\phi^{8n}}{k_1^4}\right) \quad \textrm{for } n=p
\eea 
and $\epsilon_V <1$ is only possible for $p=n$, where $V(\phi) = \frac{A^2 k_1}{M_{pl}^2} e^{k_0+\frac{\phi^{2n}}{k_1}}\left(1 - \frac{\phi^{2n}}{k_1} + \frac{\phi^{4n}}{k_1^2}\right) >0$.  In this example, $\theta$ remains a flat direction.  Unfortunately, for the well-known string theory example we considered, the deformation modulus of the deformed conifold, $n=1/3$ and $p=1$, so slow-roll is not possible.  Again, it would be extremely interesting to identify string theory constructions where $p=n$.
In the second case, the blow-up-like modulus, the slow-roll parameter is:
\bea
\epsilon_V &=& \frac{k_1}{(2\phi)^{2n}}\frac{\left(8p^2(n-p) + 12np(2n-1-2p)\frac{(2\phi)^{2n}}{k_1} + n^2(2n-3-6p)\frac{(2\phi)^{4n}}{k_1^2} - n^3\frac{(2\phi)^ {6n}}{k_1^3}\right)^2}{2n(2n-1)\left(4p^2 - 2n(6n-3-4p)\frac{(2\phi)^{2n}}{k_1} + 4n^2\frac{(2\phi)^{4n}}{k_1^2}\right)^2} \\ \nonumber
&=& \frac{4(n-p)^2}{(2n-1)2n}\frac{k_1}{(2\phi)^{2n}} + \mathcal{O}\left(\frac{(2\phi)^{2n}}{k_1}\right)
\quad \textrm{or} \quad \frac{(2\phi)^{2n}}{k_1}\frac{9}{2n(2n-1)} + \mathcal{O}\left(\frac{(2\phi)^{4n}}{k_1^2}\right) \quad \textrm{for } n=p \,.
\eea
It is straightforward to show that $\epsilon_V <1$ is only possible for $p=n$, for which \\
$V(\phi) = \frac{A^2 k_1}{2^n(n-1) ~M_{pl}^2}e^{k_0+\frac{(2\phi)^n}{k_1}}\left(n - (n-3)\frac{(2\phi)^n}{k_1} + n\frac{(2\phi)^{2n}}{k_1^2}\right) > 0$.  We must moreover have $n>1$ in order for the axion value $\theta=0$ to be metastable.

So far we have assumed that the light quintessence field starts as flat direction, and that along the runaway direction, $\phi \rightarrow \infty$ for a bulk or fibre modulus and $\phi \rightarrow 0$ for a local modulus, $W \rightarrow 0$.  We may extend the analysis to include a non-vanishing constant term, $W_0$, in the superpotential originating from the stabilisation of the heavy moduli, e.g. from fluxes.  Motivated by the simplicity of a runaway tail, we will assume in this analysis that there is no particular fine-tuning between the different ingredients.  We will also not consider the possibility of fine-tuning initial values of $\phi$ to hilltops, but focus on sourcing quintessence along the runaway tail.  The axion, $\theta$, will be set to zero. It will be helpful to introduce the following dimensionless variables
\begin{align} \label{E:xyz}
    x &= \frac{W-W_0}{W_0} &&
    y = K-k_0 \ll 1&&
    z = a\phi
\end{align}
where $y$ only applies to the local moduli ($y=\frac{\phi^{2n}}{k_1}$ for a deformation modulus and $y=\frac{(2\phi)^{2n}}{k_1}$ for a blow-up modulus and we always assume $y\ll 1$ for consistency) and $z$ only applies to the non-perturbative superpotential (in actual string theory constructions, one usually needs $z > 1$ in order to neglect higher order non-perturbative effects), while $x$ always measures the hierarchy between the terms in the superpotential ($x=\frac{Ae^{-a\phi}}{W_0}$ for the non-perturbative superpotential and $x=\frac{A\phi^p}{W_0}$ for the perturbative one, see \cite{Demirtas:2019sip} and \cite{Linde:2020mdk} for recent work on hierarchically small and large $W_0$). It is a straightforward exercise to explore the parameter space for regions that allow, simultaneously, $V(\phi) > 0$ and $\epsilon_V < 1$.  The results are summarised in Tables 1-3.  

\FloatBarrier
 \begin{table}[h]
\begin{center}
\centering
\begin{tabular}{| c | c | c | c | c |}
\hline
\multicolumn{5}{|c|}{\cellcolor[gray]{0.9}$K = -n\log(\Phi+\bar{\Phi}) \,,\quad\quad W=W_0 + Ae^{-a\Phi}$} \\
\hline\hline
\multicolumn{5}{|c|}{$V= \frac{W_0^2}{M_{pl}^2}\frac{n(n-3) (1 + x)^2 + 4 n x (1 + x) z + 4 x^2 z^2}{n2^n\phi^n}$} \\
\multicolumn{5}{|c|}{$\epsilon_V= \frac{(n (1 + x) + 2 x z)^2 (n (n - 3) (1 + x) + 4 x z (n - 1 + z))^2}{n (n (n - 3) (1 + x)^2 + 4 n x (1 + x) z + 4 x^2 z^2)^2}$} \\
\hline\hline
\multicolumn{2}{|c|}{\cellcolor[gray]{0.95} Parameters} & \cellcolor[gray]{0.95} $V\to$ & \cellcolor[gray]{0.95} $\epsilon_V\to$ & \cellcolor[gray]{0.95} $V>0 \quad \epsilon_V<1$ \\
\hline 
\multirow{2}{*}{$x\gg 1$} 
& $z \gg 1$ & $\frac{W_0^2}{M_{pl}^2}\frac{4x^2z^2}{n 2^n\phi^n}>0$ 
& $\frac{4z^2}{n}>1$ & No-go \\ \cline{2-5}
& $z \ll 1$ & $\frac{W_0^2}{M_{pl}^2}\frac{(n-3)x^2}{2^n\phi^n}$ 
& $n\geq 1$  & No-go \\ \hline
\multirow{2}{*}{$x\ll 1$} & $xz \gg 1$ & $\frac{W_0^2}{M_{pl}^2}\frac{4x^2z^2}{n 2^n\phi^n}>0$ & $\frac{4z^2}{n}>1$ & No-go \\ \cline{2-5}
& $xz \ll 1$ & $\frac{W_0^2}{M_{pl}^2}\frac{(n-3)}{2^n\phi^n}$ & $\frac{(n(n-3) + 4xz^2)^2}{n(n-3)^2}\geq 1$ & No-go \\ \hline
\multicolumn{5}{c}{} \\
\hline
\multicolumn{5}{|c|}{\cellcolor[gray]{0.9}$K = -n\log(\Phi+\bar{\Phi})\,, \quad\quad W=W_0 + A\Phi^p$} \\
\hline\hline
\multicolumn{5}{|c|}{$V= \frac{W_0^2}{M_{pl}^2}\frac{(n (x+1)-2 p x)^2-3 n (x+1)^2}{n2^n\phi^n}$} \\
\multicolumn{5}{|c|}{$\epsilon_V= \frac{(n (1 + x) - 2 p x)^2 (n (n - 3) + n (n - 3) x - 4 p x (n - p))^2}{n (4 p^2 x^2 + n^2 (1 + x)^2 - n (1 + x) (3 (1 + x) + 4 p x))^2}$} \\ 
\hline\hline
\multicolumn{2}{|c|}{\cellcolor[gray]{0.95} Parameters} & \cellcolor[gray]{0.95} $V\to$ & \cellcolor[gray]{0.95} $\epsilon_V\to$ & \cellcolor[gray]{0.95} $V>0 \quad \epsilon_V<1$ \\
\hline 
\multicolumn{2}{|c|}{$x\gg 1$} 
& $\frac{W_0^2}{M_{pl}^2}\frac{x^2 ((n-2p)^2-3n)}{n2^n\phi^n}$ & $\frac{(n-2p)^2}{n}$ & No-go \\ \hline
\multicolumn{2}{|c|}{$x\ll 1$} 
& $\frac{W_0^2}{M_{pl}^2}\frac{(n-3)}{2^n\phi^n}$ & $n$ & No-go \\ \hline
\end{tabular}
\end{center} 
\caption{Summary of interesting parameter space for string inspired supergravity models of runaway quintessence with a bulk like or fibre like modulus.  The parameters $x,y,z$ are defined and discussed in and around eq. (\ref{E:xyz}).  Note that when $W_0 = 0$, one should take the limit $W_0 x^2 \rightarrow A e^{-a\phi}$ or  $W_0 x^2 \rightarrow A \phi^p$ respectively.  In actual string compactifications, one usually requires $z = a\phi > 1$, to be able to neglect higher order non-perturbative terms.}  
\label{Table1}
\end{table}  
\FloatBarrier
 \begin{table}[h]\small
\begin{center}
\centering
\begin{tabular}{| c  c | c | c | c | c |}
\hline
\multicolumn{6}{|c|}{\cellcolor[gray]{0.9}$K = k_0 + \frac{|\Phi|^{2n}}{k_1}\,, \quad\quad W=W_0 + Ae^{-a\Phi}$} \\
\hline\hline
\multicolumn{6}{|c|}{$V= \frac{W_0^2}{M_{pl}^2}\frac{e^{k_0+y}}{n^2y}(n^2 (1 + x)^2 (y-3) y - 2 n x (1 + x) y z + x^2 z^2)$} \\
\multicolumn{6}{|c|}{$\epsilon_V= \frac{(n (1 + x) y - x z)^2 (n^2 (1 + x) (y - 2) y + 
   n x (1 - 2 y) z + (z - 1) x z)^2}{n^2 y (n^2 (1 + x)^2 (y - 3) y - 2 n x (1 + x) y z + x^2 z^2)^2}$} \\
\hline\hline
\multicolumn{3}{|c|}{\cellcolor[gray]{0.95} Parameters} & \cellcolor[gray]{0.95} $V\to$ & \cellcolor[gray]{0.95} $\epsilon_V\to$ & \cellcolor[gray]{0.95} $V>0 \quad \epsilon_V<1$\\
\hline 
\multicolumn{2}{|c|}{\multirow{2}{*}{$x\gg 1$}} & $z\gg y$ & $\frac{W_0^2}{M_{pl}^2}\frac{e^{k_0}}{n^2}x^2\big(\frac{z^2}{y} - 3n^2\big)$ & $\frac{(n-1+z)^2}{n^2(\frac{z^2}{y} - 3n^2)}\frac{z^2}{y}(\frac{z}{y})^2$ & No-go \\ \cline{3-6}
                                              & & $z\ll y$ & $-\frac{W_0^2}{M_{pl}^2}\frac{e^{k_0}}{n}x^2\left(3n + 2z\right)<0$ & $\frac{4y}{9}$ & No-go \\ \hline
\multicolumn{2}{|c|}{\multirow{2}{*}{$x\ll 1$}} & $xz\gg y$ & $\frac{W_0^2}{M_{pl}^2}\frac{e^{k_0}}{n^2}\big(\frac{(xz)^2}{y} - 3n^2\big)$ & $\frac{(n-1+z)^2}{n^2(3n^2-\frac{(xz)^2}{y})^2}\frac{(xz)^2}{y}(\frac{xz}{y})^2$ & No-go  \\ \cline{3-6}
                                              & & $xz\ll y$ & $-\frac{W_0^2}{M_{pl}^2}\frac{e^{k_0}}{n}\left(3n + 2xz\right)<0$  & $\frac{y}{9n^4}\big(\frac{xz^2}{y}-2n^2\big)^2$ & No-go \\ \hline
\end{tabular}
\\ \vspace{5mm}
\begin{tabular}{| c | c | c | c | c | c |}
\hline
\multicolumn{6}{|c|}{\cellcolor[gray]{0.9}$K = k_0 + \frac{|\Phi|^{2n}}{k_1}\,, \quad\quad W=W_0 + A\Phi^p$} \\
\hline\hline
\multicolumn{6}{|c|}{$V= \frac{W_0^2}{M_{pl}^2}\frac{e^{k_0+y}}{n^2y}((p x + n (1 + x) y)^2 - 3 n^2 (1 + x)^2 y)$} \\
\multicolumn{6}{|c|}{$\epsilon_V= \frac{(p^3 x^2 + 3 n^2 p x (1 + x) (y-1) y + n^3 (1 + x)^2 (y-2) y^2 +
   n p^2 x (y + x (3 y-1)))^2}{n^2 y (p^2 x^2 + 2 n p x (1 + x) y +
    n^2 (1 + x)^2 (y-3) y)^2}$} \\
\hline\hline
\multicolumn{3}{|c|}{\cellcolor[gray]{0.95} Parameters} & \cellcolor[gray]{0.95} $V\to$ & \cellcolor[gray]{0.95} $\epsilon_V\to$ & \cellcolor[gray]{0.95} $V>0 \quad \epsilon_V<1$ \\
\hline 
\multirow{2}{*}{$p\neq n$} & \multicolumn{2}{c|}{$x\gg 1$ } & $\frac{W_0^2}{M_{pl}^2}e^{k_0}\frac{p^2}{n^2}\frac{x^2}{y}>0$ & $\frac{(p-n)^2}{n^2}\frac{1}{y}>1$ & No-go \\ \cline{2-6}
                           & \multirow{2}{*}{$x\ll 1$} & $x\gg y$ & $\frac{W_0^2}{M_{pl}^2}\frac{p^2e^{k_0}}{n^2}\big(\frac{x^2}{y}-3\frac{n^2}{p^2}\big)$ & $\frac{(n-p)^2}{n^2\big(\frac{x^2}{y}-3\frac{n^2}{p^2}\big)^2}\frac{x^2}{y}(\frac{x}{y})^2>1$   & No-go  \\ \cline{3-6}
                           & & $x\ll y$ & $-\frac{3W_0^2e^{k_0}}{M_{pl}^2}<0$ & $\frac{4y}{9}<1$   & No-go \\ \hline
\multicolumn{3}{|c|}{\cellcolor[gray]{0.95} Parameters} & \cellcolor[gray]{0.95} $V\to$ & \cellcolor[gray]{0.95} $\epsilon_V\to$ & \cellcolor[gray]{0.95} $V>0 \quad \epsilon_V<1$ \\
\hline 
\multirow{4}{*}{$p = n$}   & \multicolumn{2}{c|}{$x\gg 1$}   & $\frac{W_0^2}{M_{pl}^2}e^{k_0}\frac{x^2}{y}>0$ & $(1+xy)^2\frac{y}{x^2}<1$    &  Yes \\ \cline{2-6}
                           & \multirow{2}{*}{$x\ll 1$} & $x^2 \gg y$ & $\frac{W_0^2}{M_{pl}^2}e^{k_0}\frac{x^2}{y}>0$ & $\frac{4y}{x^2}<1$ & Yes \\ \cline{3-6}
                           &                           & $x^2 \ll y$ & $-\frac{3W_0^2e^{k_0}}{M_{pl}^2}<0$ & $\frac{4y}{9}\big(1+\frac{x}{y}\big)^2<1$  & No-go \\ \hline
\end{tabular}
\end{center} 
\caption{Summary of interesting parameter space for string inspired supergravity models of runaway quintessence with a deformation like modulus. The parameters $x,y,z$ are defined and discussed in and around eq. (\ref{E:xyz}).  Note that when $W_0 = 0$, one should take the limit $W_0 x^2 \rightarrow A e^{-a\phi}$ or  $W_0 x^2 \rightarrow A \phi^p$ respectively.  We always assume $y \ll 1$ for consistency.  In actual string compactifications, one also usually requires $z = a\phi > 1$, to be able to neglect higher order non-perturbative terms.}
\label{Table2}
\end{table}    
\FloatBarrier
 \begin{table}[h]\small
\begin{center}
\centering
\begin{tabular}{| c  c | c | c | c | c |}
\hline
\multicolumn{6}{|c|}{\cellcolor[gray]{0.9}$K = k_0 + \frac{(\Phi+\bar{\Phi})^{2n}}{k_1}\,, \quad\quad W=W_0 + Ae^{-a\Phi}$} \\
\hline\hline
\multicolumn{6}{|c|}{$V= \frac{W_0^2}{M_{pl}^2}\frac{e^{k_0+y}}{n(2n-1)y}(2 n^2 (1 + x)^2 (y-3) y + 2 x^2 z^2 + 
  n (1 + x) y (3 - x (3 - 4 z)))$} \\
\multicolumn{6}{|c|}{$\epsilon_V= \frac{2 (2 n^3 (1 + x)^2 (y - 2) y^2 - 2 x^2 (z - 1) z^2 - 
   3 n^2 (1 + x) y (-2 x z + y (-1 + x (2z - 1))) + 
   n x z (-2 x z + y (-5 + 2 z + x (-5 + 6 z))))^2}{n (2n-1) y (2 n^2 (1 + x)^2 (y-3) y + 2 x^2 z^2 - 
   n (1 + x) y (-3 + x (4z-3)))^2}$} \\
\hline\hline
\multicolumn{3}{|c|}{\cellcolor[gray]{0.95} Parameters} & \cellcolor[gray]{0.95} $V\to$ & \cellcolor[gray]{0.95} $\epsilon_V\to$ & \cellcolor[gray]{0.95} $V>0 \quad \epsilon_V<1$ \\
\hline 
\multicolumn{2}{|c|}{\multirow{2}{*}{$x\gg 1$}} & $z \gg y$ & $\frac{W_0^2}{M_{pl}^2}\frac{2e^{k_0}}{n(2n-1)}x^2\big(\frac{z^2}{y}-\frac{3n(2n-1)}{2}\big)$ & $\frac{2(n-1+z)}{n(2n-1)\big(\frac{z^2}{y} - \frac{3n(2n-1)}{2}\big)^2}\frac{z^2}{y}(\frac{z}{y})^2$ & No-go \\ \cline{3-6}
                                              & & $z \ll y$ & $-\frac{3W_0^2e^{k_0}}{M_{pl}^2}x^2<0$ & $\frac{2 n (4 n - 3)^2}{9 (2 n - 1)^3}y<1$ & No-go  \\ \hline
\multicolumn{2}{|c|}{\multirow{2}{*}{$x\ll 1$}} & $xz \gg y$ & $\frac{W_0^2}{M_{pl}^2}\frac{2e^{k_0}}{n(2n-1)}\big(\frac{(xz)^2}{y} - \frac{3n(2n-1)}{2}\big)$ & $\frac{2(n-1+z)^2}{n (2n-1)\big(\frac{(xz)^2}{y} - \frac{3 n (2 n - 1)}{2}\big)^2}\frac{(xz)^2}{y}(\frac{xz}{y})^2$ & No-go\\ \cline{3-6}
                                                & & $xz \ll y$ & $-\frac{3W_0^2e^{k_0}}{M_{pl}^2}<0$  & $\frac{2y}{9n}\frac{(\frac{xz^2}{y}-n(4n-3))^2}{(2n-1)^3}$ & No-go \\ \hline
\end{tabular}
\\ \vspace{5mm}
\begin{tabular}{| c | c | c | c | c | c |}
\hline
\multicolumn{6}{|c|}{\cellcolor[gray]{0.9}$K = k_0 + \frac{(\Phi+\bar{\Phi})^{2n}}{k_1}\,, \quad\quad W=W_0 + A\Phi^p$} \\
\hline\hline
\multicolumn{6}{|c|}{$V= \frac{W_0^2}{M_{pl}^2}\frac{e^{k_0+y}}{n(2n-1)y}(2 p^2 x^2 + 4 n p x (1 + x) y + n (1 + x)^2 (3 + 2 n (y-3)) y)$} \\
\multicolumn{6}{|c|}{$\epsilon_V= \frac{2 (2 p^3 x^2 + 2 n^3 (1 + x)^2 (y-2) y^2 + 
   3 n^2 (1 + x) y (2 p x (y-1) + (1 + x) y) + 
   n p x (3 (1 + x) y + 2 p (y + x (3 y - 1))))^2}{n (2n - 1) y (2 p^2 x^2 + 4 n p x (1 + x) y + 
   n (1 + x)^2 (3 + 2 n (y-3)) y)^2}$} \\
\hline\hline
\multicolumn{3}{|c|}{\cellcolor[gray]{0.95} Parameters} & \cellcolor[gray]{0.95} $V\to$ & \cellcolor[gray]{0.95} $\epsilon_V\to$ & \cellcolor[gray]{0.95} $V>0 \quad \epsilon_V<1$\\
\hline 
\multirow{2}{*}{$p\neq n$} & \multicolumn{2}{c|}{$x\gg 1$ } & $\frac{W_0^2}{M_{pl}^2}\frac{2p^2e^{k_0}}{n(2n-1)}\frac{x^2}{y}>0$ & $\frac{2(n-p)^2}{n(2n-1)}\frac{1}{y}>1$ & No-go \\ \cline{2-6}
                           & \multirow{2}{*}{$x\ll 1$} & $x\gg y$ & $\frac{W_0^2}{M_{pl}^2}\frac{2p^2e^{k_0}}{n(2n-1)}\big(\frac{x^2}{y}-\frac{3n(2n-1)}{2p^2}\big)$ & $\frac{2(n-p)^2}{n(2n-1)\big(\frac{x^2}{y}-\frac{3n(2n-1)}{2p^2}\big)^2}\frac{x^2}{y}(\frac{x}{y})^2>1$   & No-go  \\ \cline{3-6}
                           & & $x\ll y$ & $-\frac{3W_0^2e^{k_0}}{M_{pl}^2}<0$ & $\frac{2n(4n-3)^2}{9(2n-1)^3}y<1$   & No-go \\ \hline
\multicolumn{3}{|c|}{\cellcolor[gray]{0.95} Parameters} & \cellcolor[gray]{0.95} $V\to$ & \cellcolor[gray]{0.95} $\epsilon_V\to$ & \cellcolor[gray]{0.95} $V>0 \quad \epsilon_V<1$ \\
\hline 
\multirow{3}{*}{$p = n$}   & \multicolumn{2}{c|}{$x\gg 1$}   & $\frac{W_0^2}{M_{pl}^2}e^{k_0}\frac{2n}{2n-1}\frac{x^2}{y}>0$ & $\frac{9y}{2n (2 n-1)} < 1$ & Yes \\ \cline{2-6}
                           & \multirow{2}{*}{$x\ll 1$} & $x^2 \gg y$ & $\frac{W_0^2}{M_{pl}^2}e^{k_0}\frac{2n}{2n-1}\frac{x^2}{y} > 0$ & $\frac{(4n-3)^2}{2n(2n-1)}\frac{y}{x^2}<1$ & Yes \\ \cline{3-6}
                           &                           & $x^2 \ll y$ & $-\frac{3W_0^2e^{k_0}}{M_{pl}^2} < 0$ & $\frac{2n}{9}\frac{(4n-3)^2}{(2n-1)^3}\big(1+\frac{x}{y}\big)^2y$ & No-go \\ \hline
\end{tabular}
\end{center} 
\caption{Summary of interesting parameter space for string inspired supergravity models of runaway quintessence with a blow-up like modulus.  The parameters $x,y,z$ are defined and discussed in and around eq. (\ref{E:xyz}).  Note that when $W_0 = 0$, one should take the limit $W_0 x^2 \rightarrow A e^{-a\phi}$ or  $W_0 x^2 \rightarrow A \phi^p$ respectively.  We always assume $y \ll 1$ for consistency.  In actual string compactifications, one also usually requires $z = a\phi > 1$, to be able to neglect higher order non-perturbative terms.}
\label{Table2}
\end{table}    
\FloatBarrier

To summarise this section, non-perturbative runaway potentials for bulk like and fibre moduli with $K = -n\ln(\Phi+\bar{\Phi})$, which contain exponentials of exponentials in the canonically normalised saxion, are too steep to source slow-roll quintessence along their tails.  One might have expected that a bulk like modulus with a perturbative runaway $W(\Phi) = A \Phi^{p}$ could lead to an exponential-like quintessence model for the canonically normalised saxion. However, we have shown that it is impossible to satisfy simultaneously $\epsilon_V<1$ and $V(\phi)>0$.  Similarly, a local modulus with $K = k_0 + \frac{(\Phi+\bar{\Phi})^{2n}}{k_1}$ or $K = k_0 + \frac{|\Phi|^{2n}}{k_1}$ with a non-perturbative runaway would have an exponential envelope in the scalar potential for the canonically normalised saxion, but it turns out to be impossible to realise slow-roll quintessence.  A local modulus with a perturbative runaway allows power-law slow-roll quintessence within supergravity only in very special cases, where the leading power in the superpotential, $p$, is equal to $n$ in the leading power in the K\"ahler potential.  It would be very interesting to find concrete string theory realisations of this scenario.

\section{Thermal Dark Energy}

Given the difficulties encountered in building de Sitter vacua and quintessence in string theory, and the several upcoming observational probes into Dark Energy, it is extremely interesting to consider well-motivated alternative Dark Energy scenarios in string theory.  In the remainder of this talk we will review the proposal \cite{Hardy:2019apu}.

Although reliable metastable de Sitter vacua in string theory are hard to find, we know that unstable de Sitter vacua do exist.  Consider e.g. the simplest example of a bulk modulus $K=-\ln(\Phi+\bar{\Phi})$ and a non-perturbative superpotential $W= A e^{-a \Phi}$, illustrated in Figure \ref{F:pot}.  Moreover, there are strong reasons to believe that unstable de Sitter vacua exist in Nature, most notably from the Mexican hat potential of the Higgs field.  Reflecting on the Higgs field, it is interesting to note moreover that high temperature effects transform the Higg's unstable de Sitter vacuum to a metastable de Sitter one, in which the electroweak symmetry is restored.  So it would also seem that metastable de Sitter vacua produced via high temperature effects exist in Nature.  In the history of our Universe, this positive potential energy density did not lead to an accelerated expansion because it was dominated by the radiation energy density.  However, as we will now show, in string theory there may very well be light hidden sectors, in which thermal effects generate a metastable de Sitter minimum that temporarily dominates the Universe and drives an accelerated expansion.

Consider a light hidden sector that is in internal thermal equilibrium in the present day Universe, that is interaction rates in the hidden sector are larger than the Hubble expansion rate, $\Gamma_I > H$, with some hidden temperature $T_h$.  Suppose that it includes a scalar field, $\phi$, whose zero temperature potential energy functional, $V(\phi)$, stabilises it to some vacuum expectation value, $\langle \phi \rangle = \phi_1$.  For example, we may consider a Higgs-like quartic potential:
\be
V(\phi) = \lambda \phi^4 - \frac{m_\phi^2}{2}\phi^2 + C \label{E:zeropot}
\ee
for which $\phi_1 = m_\phi/\left(2\sqrt{\lambda}\right)$.  We will fix the constant $C=m_\phi^4/\left(16\lambda\right)$ to give $V(\phi_1)=0$, though this condition can be relaxed.  Suppose moreover that $\phi$ has Higgs-like interactions with other hidden sector states, e.g. hidden fermions with interactions $y_i \phi \bar{\psi}^i \psi^i$ or hidden scalars with interactions $\lambda_a \phi^2 \chi^a \chi^a$.  When $\phi$ acquires a vacuum expectation value, these hidden fermions or scalars acquire masses.

The plasma of light hidden sector particles interacts with the homogeneous scalar field background, $\langle \phi \rangle = \phi_c$, whilst this background also determines the masses and interactions of the particles in the plasma.  This leads to a thermal effective potential for $\phi$, which at one loop takes the form (see e.g. \cite{Bellac:2011kqa}):
\be
V(\phi_c, T_h) = V_0(\phi_c) + \frac{T_h^4}{2\pi^2}\left(-\sum_{\psi^i}n_{\psi^i}~ J_F\left(\frac{m^2_{\psi^i}(\phi_c)}{T_h^2}\right) + \sum_{\chi^a} n_{\chi^a}~ J_B\left(\frac{M^2_{\chi^a}(\phi_c)}{T_h^2}\right)\right)
\ee
where $n_{\psi^i}$ and $n_{\chi^a}$ are the number of degrees of freedom in the fermion $\psi^i$ and scalar $\chi^a$, respectively; the mass-squareds of the hidden sector fermions and scalars are, respectively, $m_{\psi^i}^2(\phi_c) = y_i^2 \phi_c^2$ and $M^2_{\chi^a}(\phi_c) = \lambda_a \phi_c^2$; and $J_{F/B}$ are the fermionic/bosonic thermal functions, whose high temperature limits are, respectively:
\bea
J_F(x^2) &=& \frac{7\pi^4}{360} - \frac{\pi^2}{24}x^2 + \mathcal{O}(x^3) \quad \textrm{when }|x| \ll 1 \nonumber \\
J_B(x^2) &=& -\frac{\pi^4}{45} + \frac{\pi^2}{12}x^2 + \mathcal{O}(x^3) \quad \textrm{when }|x| \ll 1
\eea
(in the low temperature limit, the thermal functions are exponentially suppressed).  Therefore, at temperatures much higher than the masses in the thermal bath, $T_h \gg m_{\psi}(\phi_c),M_{\chi}(\phi_c)$, the thermally corrected scalar potential for $\phi$ takes the form:
\be
V(\phi, T_h) = \lambda \phi^4 - \frac{m_\phi^2}{2}\phi^2 + \frac{m_\phi^4}{16\lambda}- a T_h^4 + b T_h^2 \phi^2 \,, \label{E:thermalpot}
\ee
where $a$ and $b$ can be inferred from the expressions above, and depend on $n_{\psi^i}$, $n_{\chi^a}$, $y_i$ and $\lambda_a$.  

\begin{figure}[t!]
\begin{center}
\includegraphics[width=0.4\textwidth]{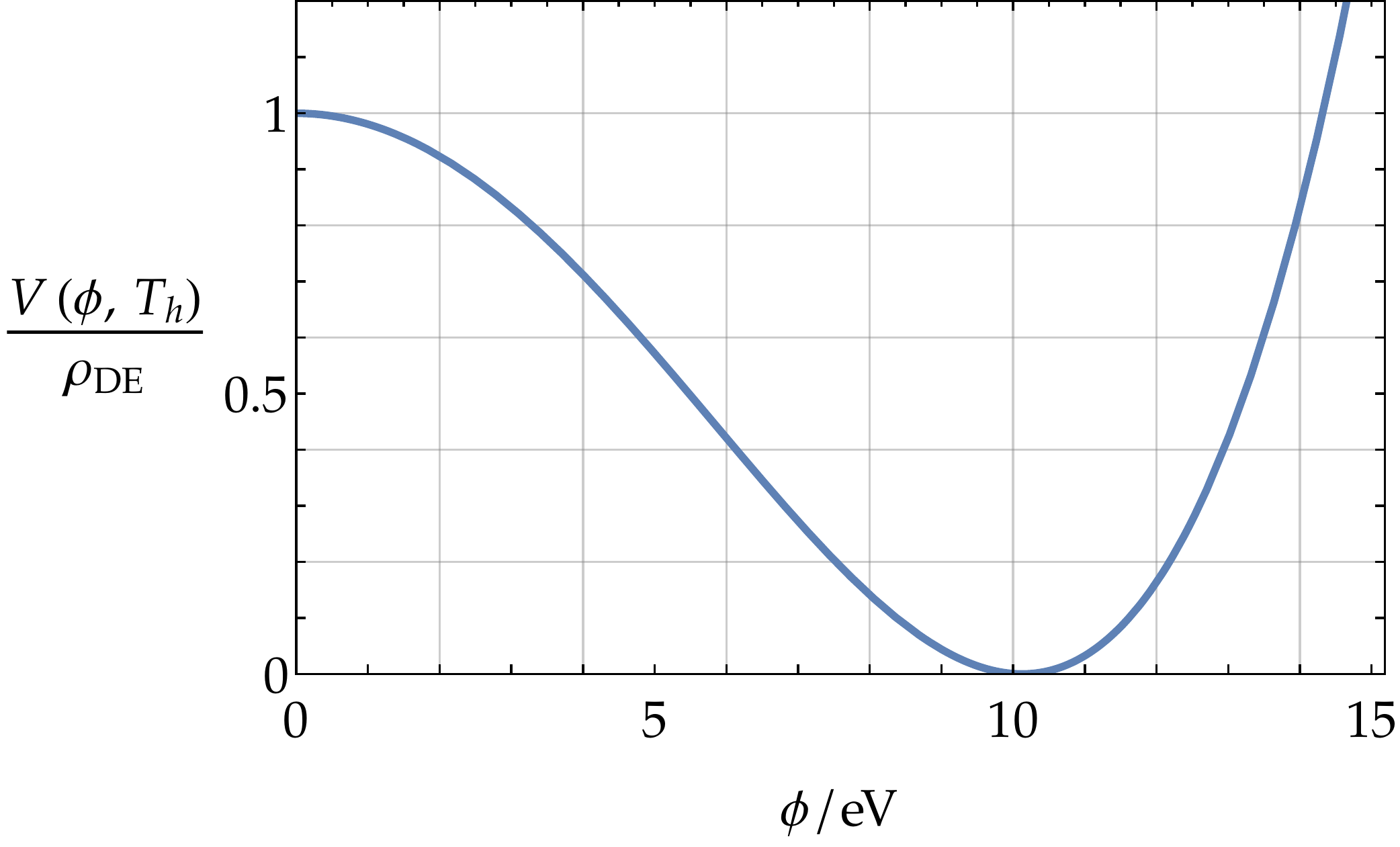} \hspace{1cm} \includegraphics[width=0.4\textwidth]{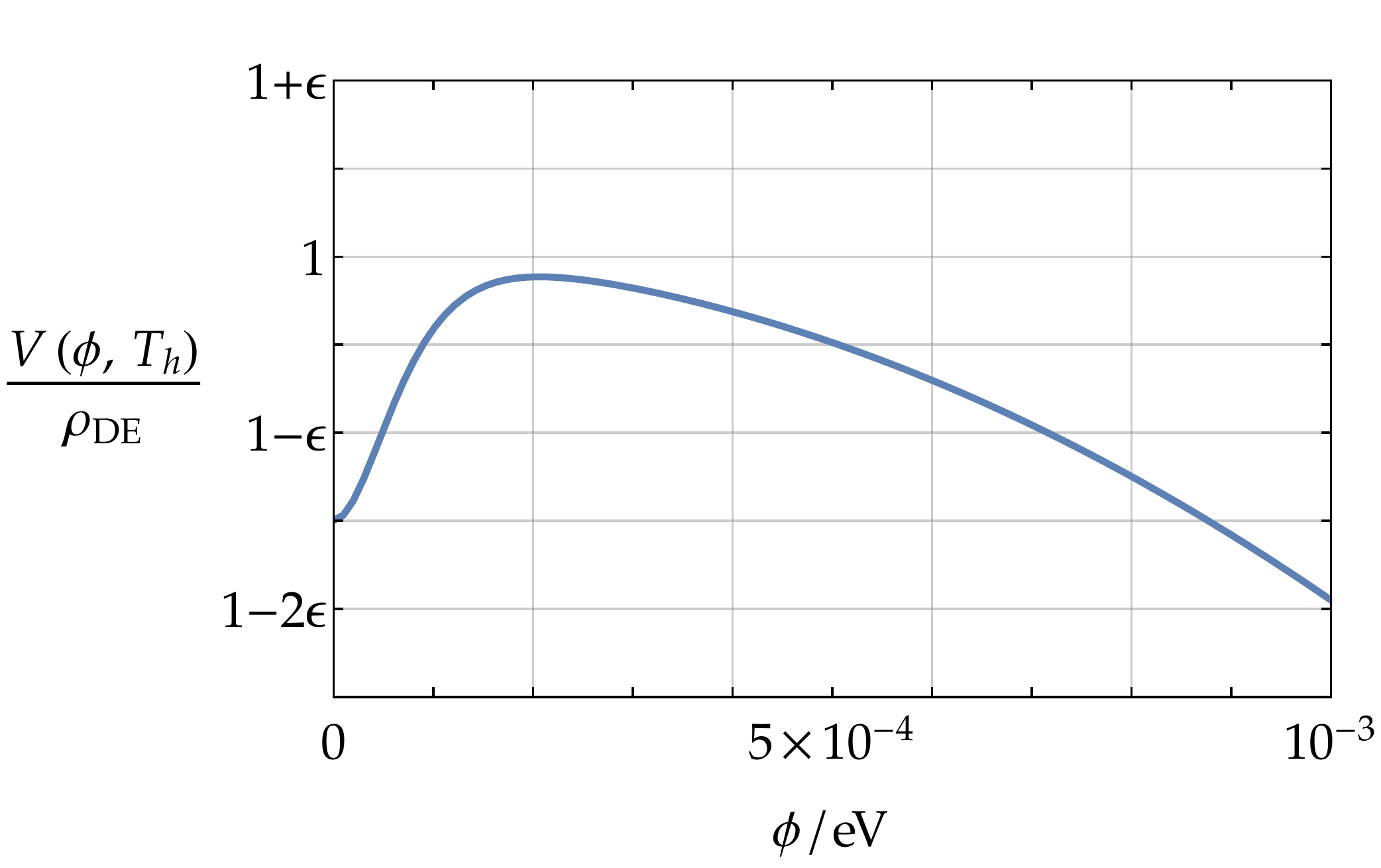}
\end{center}
\caption{Finite temperature scalar potential (\ref{E:thermalpot}) for a model with $\phi$ coupled to a single Dirac fermion with Yukawa coupling $y=1$, and $m_\phi = 10^{-6}$eV and $\lambda = 2.44 \times 10^{-15}$.  The hidden sector temperature is fixed to $T_h = 10^{-4.5}$eV.  The potential is symmetric about $\phi=0$. The vertical axis has been normalised to the present day dark energy density, $\rho_{DE} = (2.3\textrm{meV})^4$.  In the right panel a close up is given, with $\epsilon = 10^{-8}$.  Figure reproduced from \cite{Hardy:2019apu}.\label{F:thermalpot}}
\end{figure}

For sufficiently high temperatures:
\be
T_h > \frac{m_\phi}{\sqrt{2b}} \,, \label{E:newmin}
\ee
 the minimum of the scalar potential (\ref{E:thermalpot}) is shifted from $\phi=\phi_1$ to $\phi=0$.  Since at $\phi=0$, states in the thermal bath are massless, the high temperature approximation used in eqn. (\ref{E:thermalpot}) is automatically valid.  At temperatures so high that eqn. (\ref{E:thermalpot}) is valid even at $\phi=\phi_1$, $\phi=0$ becomes a global minimum.  As the temperature falls, the high temperature approximation may break down at larger $\phi$, where hidden fermions and scalars are heavy.  Then the zero temperature potential dominates in this part of the field space, and there is a minimum close to $\phi=\phi_1$, which is deeper than the minimum at $\phi=0$ if $T^4 \ll m_\phi^4/\lambda$.  See Figure \ref{F:thermalpot} for an illustrative example of the finite temperature scalar potential (\ref{E:thermalpot}).

As we have seen, for sufficiently high temperatures, $\phi$ is trapped in a minimum at $\phi=0$.  Because of the shifted vacuum expectation value of $\phi$, it carries a non-trivial potential energy\footnote{Note that the finite temperature terms in the potential do not directly contribute to the Einstein's equations, see e.g. \cite{Chung:2011it}.} that contributes to the dark energy density:
\be
\rho_{DE} = V(\phi=0) = \frac{m_\phi^4}{16\lambda} \,.\label{E:rhoDE}
\ee
Intriguingly, the scale of the observed Dark Energy is roughly the same order of magnitude as the  meV upper bound from fifth force constraints for light scalars with Planckian couplings to the visible sector.  However, there are other constraints on the model that narrow the window of the parameter space.

To source an accelerated expansion, $\rho_{DE}$ must dominate over the hidden sector radiation density:
\be
\frac{m_\phi^4}{16 \lambda} > \frac{\pi^2 g_h {T_h^0}^4}{30} \,, \label{E:rhorad}
\ee
where $g_h$ is the number of hidden sector relativistic degrees of freedom and $T_h^0$ is the hidden sector temperature today.  Note also that observational bounds on the total effective number of relativistic degrees of freedom at early times imply that $T_h^0 < T_v^0$, where the temperature of the visible sector CMB photons today is $T_v^0 \sim 0.24$ meV.  Put together with the condition that finite temperature effects are sufficient to produce a new minimum at $\phi=0$ (\ref{E:newmin}) implies that:
\be
m_\phi \ll T_h \quad \textrm{and} \quad \lambda \ll 1 \,.
\ee
For example, with a single Dirac fermion with Yukawa coupling $y=1$, and $m_\phi = 10^{-6}$eV, $\lambda = 2.44 \times 10^{-15}$ and $T_h = 10^{-4.5}$eV, the potential energy of the field $\phi$ that is induced by finite temperature effects matches the Dark Energy density and dominates our Universe today.  The Dark Energy epoch ends when $T_h \sim m_{\phi}$, with the onset of a first order phase transition towards the true vacuum, and conversion to hidden sector radiation, matter and gravitational waves. For the parameters just mentioned, the total number of efolds of Dark Energy domination can be computed to be $N_{DE}=4.1$.

\begin{figure}[t!]
\begin{center}
\includegraphics[width=0.4\textwidth]{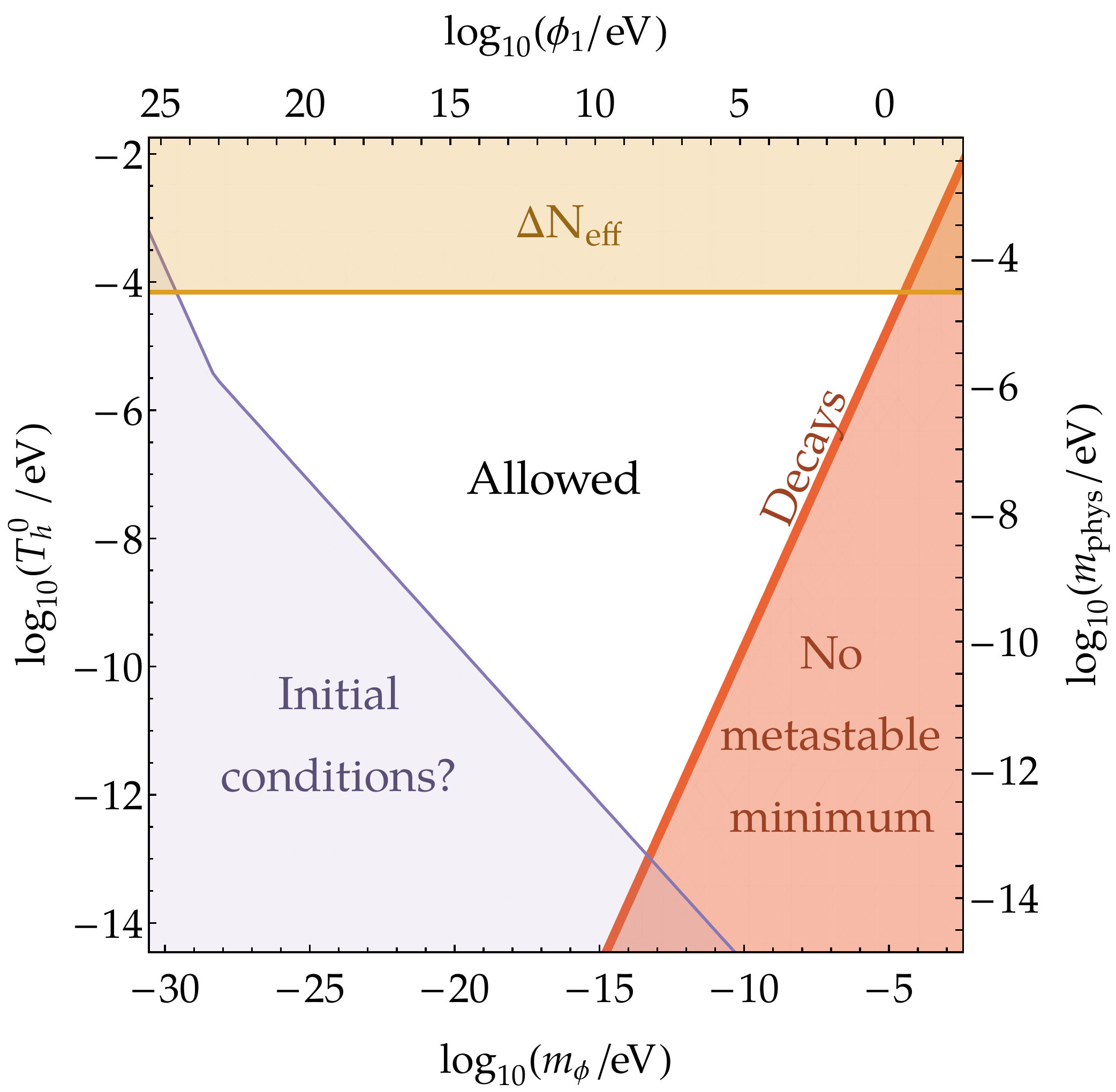}
\end{center}
\caption{Allowed parameter space for Thermal Dark Energy with zero temperature scalar potential (\ref{E:zeropot}) and a single Dirac fermion with $y=1$. For details see text.  Figure reproduced from \cite{Hardy:2019apu}. \label{F:paramspace}}
\end{figure}
Figure \ref{F:paramspace} shows the parameter space for which a scalar with zero temperature potential (\ref{E:zeropot}), coupled to a single Dirac fermion with $y=1$, can give rise to a Thermal Dark Energy that moreover matches the observed Dark Energy.  Note that the Lagrangian mass parameter, $m_\phi$ does not correspond to the physical mass of the scalar around the metastable minimum in the present day Universe, which is sourced by thermal effects:
\be
m_{phys}^2 = 2 b {T_h^0}^2 \,.
\ee
Also shown in the figure are constraints on $T_h^0 \lesssim 0.3 T_v^0$ arising from the effective number of neutrino species, $\Delta N_{eff}$; constraints from requiring that the metastable minimum is sufficiently long-lived, that is the rate of bubble-nucleation to the true vacuum is sufficiently suppressed $\Gamma_{nucl}\ll H_0^4$; constraints from requiring that the metastable minimum is a global minimum in the early Universe, thus explaining how the scalar field starts in what is today a very shallow local minimum.  

For the simplest models, the time dependence of Dark Energy is negligible, and the model predicts an equation of state $w_{DE}=-1$.  Portal interactions between visible and hidden sectors are constrained by fifth forces and the requirement that the hidden sector stays cool to keep $\Delta N_{eff}$ small.  Larger portal couplings increase the chances of observing fifth forces in future experiments, however, they also increase the amount of fine-tuning necessary in the model.  Note that the amount of sequestering needed for a scalar of mass around $m \sim 10^{-6}$ eV is much easier to achieve than what is required for a quintessence field, with $m \sim 10^{-33}$ eV.  As well as deviations from the Standard Model prediction for $\Delta N_{eff}$, possibly the most promising observational signatures for Thermal Dark Energy are remnants from earlier Thermal Dark Energy\footnote{Thermal Inflation \cite{Lyth:1995ka} may be one such epoch sourced by the visible sector itself at scales $\sim$TeV.} epochs, where hidden sectors associated with distinct mass scales provide a setup for the Early Dark Energy scenario \cite{Karwal:2016vyq}.  Although model dependent, the decay of early Thermal Dark Energy components could lead to stable hidden sector relics and a stochastic background of gravitational waves.  Early Thermal Dark Energies may also help resolve the discrepancy between cosmological and astrophysical measurements of today's Hubble parameter, $H_0$ (for a review see \cite{Verde:2019ivm}).

The requirement of a small dimensionless parameter in order for the Thermal Dark Energy to dominate over the total radiation density is seen for other forms of the zero-temperature scalar potential.  Indeed, the absence of such a hierarchy for the Standard Model Higgs is why there is no period of Thermal Dark Energy domination before electroweak symmetry breaking.  Attractively, the hierarchy required in Thermal Dark Energy can be made technically natural, by embedding it in a mildly sequestered supersymmetric hidden sector.
 
\section{Outlook}
Our current knowledge of string theory requires us to work always within the weak coupling and large volume perturbative limits.  The Dine-Seiberg runaway argument and several no-go theorems then indicate that moduli stabilisation in a de Sitter vacuum will be at the limits of theoretical control, requiring a delicate interplay between diverse string theoretic ingredients.  We have focussed, therefore, on two alternative scenarios for Dark Energy: runaway quintessence (reviewing \cite{Olguin-Tejo:2018pfq} and presenting some further results) and Thermal Dark Energy \cite{Hardy:2019apu}.

Runaway moduli are ubiquitous in string theory constructions, and a very simple explanation for Dark Energy would be a runaway string modulus, where the saxion is frozen or slowly rolling down its runaway tail.  We have considered several simple, single-field, supergravity runaway directions motivated by string theory, where the leading superpotential goes as $W = W_0+A e^{-a \Phi}$ or $W = W_0+A \Phi^p$ (including the case where the $\Phi$ starts off as a flat direction with $W_0=0$).  Bulk and fibre moduli where $K = -n\ln(\Phi+\bar{\Phi})$, cannot sustain slow-roll quintessence, as it is impossible to satisfy simultaneously $V>0$ and $\epsilon_V<1$.  A local deformation-like modulus, where $K = k_0 + \frac{|\Phi|^{2n}}{k_1}$ can achieve $V>0$ and $\epsilon_V<1$ only for the perturbative superpotential with $p=n$, with a minimum for the axion if $n>1$. A local blow-up-like modulus, where $K = k_0 + \frac{(\Phi+\bar{\Phi})^{2n}}{k_1}$, can similarly achieve $V>0$ and $\epsilon_V<1$ only for the perturbative superpotential with $p=n$, where now the axion remains a flat direction, possibly stabilised by subleading effects.  Unfortunately, the well-known deformation modulus for the deformed conifold does not satisfy the required condition.  It would be very interesting to identify concrete string moduli with runaway superpotentials that does fulfill it.

Thermal Dark Energy is motivated by the prevalence of hidden sectors in string theory, and the challenges in obtaining both de Sitter vacua and slow-roll directions in explicit string constructions.  Finite-temperature effects in a light hidden sector can explain Dark Energy, by holding a light scalar field away from the minimum of the zero-temperature potential.  The resulting potential dark energy has equation of state parameter $w=-1$, and yet is consistent with the Swampland Conjectures.  In order that the dark energy dominates over dark radiation, some hierarchy in the model parameters is necessary.  For scalar masses $m_\phi \lesssim \mu$eV, which could be technically natural via sequestering, there are large regions of viable parameter space consistent with observations.  Thermal Dark Energy is potentially observable via fifth forces and $\Delta N_{eff}$, and earlier Thermal Dark Energy epochs could explain the $H_0$ tension and potentially leave interesting gravitational wave signatures.  It would be very interesting to explore in more detail these signatures, as well as to embed the scenario in an ultraviolet complete theory like string theory.

Finally, in view of the great mystery that Dark Energy presents, it remains vital to conceive of other well-motivated candidates for its origin, and their possible signatures for the impending Dark Energy observations.

\acknowledgments{We warmly acknowledge and thank Ed Hardy, Yessenia Olgu\'in-Trejo and Gianmassimo Tasinato with whom some of this work was done.  S. L. P thanks the organisers and participants of the Corfu Summer Institute 2019, for the stimulating and enjoyable workshop.  D. C. is supported by a CONACyT Mexico grant and Beca-Mixta CONACyT. D. C. thanks the Theoretical Physics group at University of Liverpool for an amazing hospitality and support.  I. Z. is partially supported by STFC, grant ST/P00055X/1.}


\begin{thebibliography}{99}

\bibitem{Martin:2012bt}
J.~Martin,
Comptes Rendus Physique \textbf{13} (2012), 566-665
doi:10.1016/j.crhy.2012.04.008
[arXiv:1205.3365 [astro-ph.CO]].

\bibitem{Dine:1985he}
M.~Dine and N.~Seiberg,
Phys. Lett. B \textbf{162} (1985), 299-302
doi:10.1016/0370-2693(85)90927-X

\bibitem{Gibbons:1984kp}
G.~Gibbons,
GIFT Seminar on supersymmetry, supergravity and related topics, edited by F. del Aguila, J. de Ascarraga and L. Ibanez, World Scientific (1984)
Print-85-0061 (CAMBRIDGE).

\bibitem{Maldacena:2000mw}
J.~M.~Maldacena and C.~Nunez,
Int. J. Mod. Phys. A \textbf{16} (2001), 822-855
doi:10.1142/S0217751X01003937
[arXiv:hep-th/0007018 [hep-th]].

\bibitem{Kutasov:2015eba}
D.~Kutasov, T.~Maxfield, I.~Melnikov and S.~Sethi,
Phys. Rev. Lett. \textbf{115} (2015) no.7, 071305
doi:10.1103/PhysRevLett.115.071305
[arXiv:1504.00056 [hep-th]].

\bibitem{Giddings:2001yu}
S.~B.~Giddings, S.~Kachru and J.~Polchinski,
Phys. Rev. D \textbf{66} (2002), 106006
doi:10.1103/PhysRevD.66.106006
[arXiv:hep-th/0105097 [hep-th]].

\bibitem{Kachru:2003aw}
S.~Kachru, R.~Kallosh, A.~D.~Linde and S.~P.~Trivedi,
Phys. Rev. D \textbf{68} (2003), 046005
doi:10.1103/PhysRevD.68.046005
[arXiv:hep-th/0301240 [hep-th]].

\bibitem{Susskind:2003kw}
L.~Susskind,
[arXiv:hep-th/0302219 [hep-th]].

\bibitem{Weinberg:1988cp}
S.~Weinberg,
Rev. Mod. Phys. \textbf{61} (1989), 1-23
doi:10.1103/RevModPhys.61.1

\bibitem{Danielsson:2018ztv}
U.~H.~Danielsson and T.~Van Riet,
Int. J. Mod. Phys. D \textbf{27} (2018) no.12, 1830007
doi:10.1142/S0218271818300070
[arXiv:1804.01120 [hep-th]].

\bibitem{Cicoli:2018kdo}
M.~Cicoli, S.~De Alwis, A.~Maharana, F.~Muia and F.~Quevedo,
Fortsch. Phys. \textbf{67} (2019) no.1-2, 1800079
doi:10.1002/prop.201800079
[arXiv:1808.08967 [hep-th]].

\bibitem{Obied:2018sgi}
G.~Obied, H.~Ooguri, L.~Spodyneiko and C.~Vafa,
[arXiv:1806.08362 [hep-th]].

\bibitem{Garg:2018reu}
S.~K.~Garg and C.~Krishnan,
JHEP \textbf{11} (2019), 075
doi:10.1007/JHEP11(2019)075
[arXiv:1807.05193 [hep-th]].

\bibitem{Ooguri:2018wrx}
H.~Ooguri, E.~Palti, G.~Shiu and C.~Vafa,
Phys. Lett. B \textbf{788} (2019), 180-184
doi:10.1016/j.physletb.2018.11.018
[arXiv:1810.05506 [hep-th]].

\bibitem{Bedroya:2019snp}
A.~Bedroya and C.~Vafa,
[arXiv:1909.11063 [hep-th]].

\bibitem{Witten:2001kn}
E.~Witten,
[arXiv:hep-th/0106109 [hep-th]].

\bibitem{Banks:2012hx}
T.~Banks,
[arXiv:1208.5715 [hep-th]].

\bibitem{Maltz:2016iaw}
J.~Maltz and L.~Susskind,
Phys. Rev. Lett. \textbf{118} (2017) no.10, 101602
doi:10.1103/PhysRevLett.118.101602
[arXiv:1611.00360 [hep-th]].

\bibitem{Dvali:2018jhn}
G.~Dvali, C.~Gomez and S.~Zell,
Fortsch. Phys. \textbf{67} (2019) no.1-2, 1800094
doi:10.1002/prop.201800094
[arXiv:1810.11002 [hep-th]].

\bibitem{Tsujikawa:2013fta}
S.~Tsujikawa,
Class. Quant. Grav. \textbf{30} (2013), 214003
doi:10.1088/0264-9381/30/21/214003
[arXiv:1304.1961 [gr-qc]].

\bibitem{Hebecker:2019csg}
A.~Hebecker, T.~Skrzypek and M.~Wittner,
JHEP \textbf{11} (2019), 134
doi:10.1007/JHEP11(2019)134
[arXiv:1909.08625 [hep-th]].

\bibitem{Freese:1990rb}
K.~Freese, J.~A.~Frieman and A.~V.~Olinto,
Phys. Rev. Lett. \textbf{65} (1990), 3233-3236
doi:10.1103/PhysRevLett.65.3233

\bibitem{Svrcek:2006hf}
P.~Svrcek,
[arXiv:hep-th/0607086 [hep-th]].

\bibitem{Banks:2003sx}
T.~Banks, M.~Dine, P.~J.~Fox and E.~Gorbatov,
JCAP \textbf{06} (2003), 001
doi:10.1088/1475-7516/2003/06/001
[arXiv:hep-th/0303252 [hep-th]].

\bibitem{ArkaniHamed:2006dz}
N.~Arkani-Hamed, L.~Motl, A.~Nicolis and C.~Vafa,
JHEP \textbf{06} (2007), 060
doi:10.1088/1126-6708/2007/06/060
[arXiv:hep-th/0601001 [hep-th]].

\bibitem{Kim:2004rp}
J.~E.~Kim, H.~P.~Nilles and M.~Peloso,
JCAP \textbf{01} (2005), 005
doi:10.1088/1475-7516/2005/01/005
[arXiv:hep-ph/0409138 [hep-ph]].

\bibitem{McAllister:2014mpa}
L.~McAllister, E.~Silverstein, A.~Westphal and T.~Wrase,
JHEP \textbf{09} (2014), 123
doi:10.1007/JHEP09(2014)123
[arXiv:1405.3652 [hep-th]].

\bibitem{Parameswaran:2016qqq}
S.~Parameswaran, G.~Tasinato and I.~Zavala,
JCAP \textbf{04} (2016), 008
doi:10.1088/1475-7516/2016/04/008
[arXiv:1602.02812 [astro-ph.CO]].

\bibitem{Cicoli:2012tz}
M.~Cicoli, F.~G.~Pedro and G.~Tasinato,
JCAP \textbf{07} (2012), 044
doi:10.1088/1475-7516/2012/07/044
[arXiv:1203.6655 [hep-th]].


\bibitem{Olguin-Tejo:2018pfq}
Y.~Olguin-Trejo, S.~L.~Parameswaran, G.~Tasinato and I.~Zavala,
JCAP \textbf{01} (2019), 031
doi:10.1088/1475-7516/2019/01/031
[arXiv:1810.08634 [hep-th]].


\bibitem{Hardy:2019apu}
E.~Hardy and S.~Parameswaran,
Phys. Rev. D \textbf{101} (2020) no.2, 023503
doi:10.1103/PhysRevD.101.023503
[arXiv:1907.10141 [hep-th]].

\bibitem{Hertzberg:2007wc}
M.~P.~Hertzberg, S.~Kachru, W.~Taylor and M.~Tegmark,
JHEP \textbf{12} (2007), 095
doi:10.1088/1126-6708/2007/12/095
[arXiv:0711.2512 [hep-th]].

\bibitem{Wrase:2010ew}
T.~Wrase and M.~Zagermann,
Fortsch. Phys. \textbf{58} (2010), 906-910
doi:10.1002/prop.201000053
[arXiv:1003.0029 [hep-th]].

\bibitem{Copeland:1997et}
E.~J.~Copeland, A.~R.~Liddle and D.~Wands,
Phys. Rev. D \textbf{57} (1998), 4686-4690
doi:10.1103/PhysRevD.57.4686
[arXiv:gr-qc/9711068 [gr-qc]].

\bibitem{Copeland:2006wr}
E.~J.~Copeland, M.~Sami and S.~Tsujikawa,
Int. J. Mod. Phys. D \textbf{15} (2006), 1753-1936
doi:10.1142/S021827180600942X
[arXiv:hep-th/0603057 [hep-th]].



\bibitem{Agrawal:2018own}
P.~Agrawal, G.~Obied, P.~J.~Steinhardt and C.~Vafa,
Phys. Lett. B \textbf{784} (2018), 271-276
doi:10.1016/j.physletb.2018.07.040
[arXiv:1806.09718 [hep-th]].

\bibitem{Caviezel:2008tf}
C.~Caviezel, P.~Koerber, S.~Kors, D.~Lust, T.~Wrase and M.~Zagermann,
JHEP \textbf{04} (2009), 010
doi:10.1088/1126-6708/2009/04/010
[arXiv:0812.3551 [hep-th]].

\bibitem{Flauger:2008ad}
R.~Flauger, S.~Paban, D.~Robbins and T.~Wrase,
Phys. Rev. D \textbf{79} (2009), 086011
doi:10.1103/PhysRevD.79.086011
[arXiv:0812.3886 [hep-th]].

\bibitem{Roupec:2018mbn}
C.~Roupec and T.~Wrase,
Fortsch. Phys. \textbf{67} (2019) no.1-2, 1800082
doi:10.1002/prop.201800082
[arXiv:1807.09538 [hep-th]].

\bibitem{Parameswaran:2010ec}
S.~L.~Parameswaran, S.~Ramos-Sanchez and I.~Zavala,
JHEP \textbf{01} (2011), 071
doi:10.1007/JHEP01(2011)071
[arXiv:1009.3931 [hep-th]].

\bibitem{Anderson:2011cza}
L.~B.~Anderson, J.~Gray, A.~Lukas and B.~Ovrut,
Phys. Rev. D \textbf{83} (2011), 106011
doi:10.1103/PhysRevD.83.106011
[arXiv:1102.0011 [hep-th]].

\bibitem{Cicoli:2013rwa}
M.~Cicoli, S.~de Alwis and A.~Westphal,
JHEP \textbf{10} (2013), 199
doi:10.1007/JHEP10(2013)199
[arXiv:1304.1809 [hep-th]].

\bibitem{Danielsson:2012et}
U.~H.~Danielsson, G.~Shiu, T.~Van Riet and T.~Wrase,
JHEP \textbf{03} (2013), 138
doi:10.1007/JHEP03(2013)138
[arXiv:1212.5178 [hep-th]].

\bibitem{Cicoli:2009fibre}
M.~Cicoli, F.~Quevedo and C. P. Burgess,
Journal of Cosmology and Astroparticle Physics, 2009(03), 013
[arXiv:0808.0691v3 [hep-th]].

\bibitem{Gonzalo:2018guu}
E.~Gonzalo, L.~E.~Ibáñez and Á.~M.~Uranga,
JHEP \textbf{05} (2019), 105
doi:10.1007/JHEP05(2019)105
[arXiv:1812.06520 [hep-th]].

\bibitem{Balasubramanian:2005zx}
V.~Balasubramanian, P.~Berglund, J.~P.~Conlon and F.~Quevedo,
JHEP \textbf{03} (2005), 007
doi:10.1088/1126-6708/2005/03/007
[arXiv:hep-th/0502058 [hep-th]].

\bibitem{Westphal:2006tn}
A.~Westphal,
JHEP \textbf{03} (2007), 102
doi:10.1088/1126-6708/2007/03/102
[arXiv:hep-th/0611332 [hep-th]].

\bibitem{Cicoli:2015ylx}
M.~Cicoli, F.~Quevedo and R.~Valandro,
JHEP \textbf{03} (2016), 141
doi:10.1007/JHEP03(2016)141
[arXiv:1512.04558 [hep-th]].


\bibitem{Blaback:2013qza}
J.~Blab\"ack, D.~Roest and I.~Zavala,
Phys. Rev. D \textbf{90} (2014) no.2, 024065
doi:10.1103/PhysRevD.90.024065
[arXiv:1312.5328 [hep-th]].


\bibitem{Bena:2018fqc}
I.~Bena, E.~Dudas, M.~Graña and S.~Lüst,
Fortsch. Phys. \textbf{67} (2019) no.1-2, 1800100
doi:10.1002/prop.201800100
[arXiv:1809.06861 [hep-th]].

\bibitem{Sethi:2017phn}
S.~Sethi,
JHEP \textbf{10} (2018), 022
doi:10.1007/JHEP10(2018)022
[arXiv:1709.03554 [hep-th]].



\bibitem{Blaback:2013ht}
J.~Blåbäck, U.~Danielsson and G.~Dibitetto,
JHEP \textbf{08} (2013), 054
doi:10.1007/JHEP08(2013)054
[arXiv:1301.7073 [hep-th]].

\bibitem{Kachru:2019dvo}
S.~Kachru, M.~Kim, L.~McAllister and M.~Zimet,
[arXiv:1908.04788 [hep-th]].

\bibitem{Gautason:2019jwq}
F.~Gautason, V.~Van Hemelryck, T.~Van Riet and G.~Venken,
[arXiv:1902.01415 [hep-th]].

\bibitem{Grana:2020hyu}
M.~Graña, N.~Kovensky and A.~Retolaza,
[arXiv:2002.01481 [hep-th]].

\bibitem{Cordova:2018dbb}
C.~Córdova, G.~B.~De Luca and A.~Tomasiello,
Phys. Rev. Lett. \textbf{122} (2019) no.9, 091601
doi:10.1103/PhysRevLett.122.091601
[arXiv:1812.04147 [hep-th]].

\bibitem{Cordova:2019dbb}
C.~Córdova, G.~B.~De Luca and A.~Tomasiello,
[arXiv:1911.04498 [hep-th]].

\bibitem{Dasgupta:2019rwt}
K.~Dasgupta, M.~Emelin, M.~Mehedi Faruk and R.~Tatar,
[arXiv:1911.12382 [hep-th]].

\bibitem{Berg:2010ha}
M.~Berg, D.~Marsh, L.~McAllister and E.~Pajer,
JHEP \textbf{06} (2011), 134
doi:10.1007/JHEP06(2011)134
[arXiv:1012.1858 [hep-th]].

\bibitem{Aparicio:2014wxa}
L.~Aparicio, M.~Cicoli, S.~Krippendorf, A.~Maharana, F.~Muia and F.~Quevedo,
JHEP \textbf{11} (2014), 071
doi:10.1007/JHEP11(2014)071
[arXiv:1409.1931 [hep-th]].

\bibitem{Acharya:2018deu}
B.~S.~Acharya, A.~Maharana and F.~Muia,
JHEP \textbf{03} (2019), 048
doi:10.1007/JHEP03(2019)048
[arXiv:1811.10633 [hep-th]].

\bibitem{Dine:1986vd}
M.~Dine and N.~Seiberg,
Phys. Rev. Lett. \textbf{57} (1986), 2625
doi:10.1103/PhysRevLett.57.2625

\bibitem{Burgess:2005jx}
C.~Burgess, C.~Escoda and F.~Quevedo,
JHEP \textbf{06} (2006), 044
doi:10.1088/1126-6708/2006/06/044
[arXiv:hep-th/0510213 [hep-th]].

\bibitem{GarciadelMoral:2017vnz}
M.~P.~Garcia del Moral, S.~Parameswaran, N.~Quiroz and I.~Zavala,
JHEP \textbf{10} (2017), 185
doi:10.1007/JHEP10(2017)185
[arXiv:1707.07059 [hep-th]].

\bibitem{Cicoli:2008va}
M.~Cicoli, J.~P.~Conlon and F.~Quevedo,
JHEP \textbf{10} (2008), 105
doi:10.1088/1126-6708/2008/10/105
[arXiv:0805.1029 [hep-th]].

\bibitem{Douglas:2007tu}
M.~R.~Douglas, J.~Shelton and G.~Torroba,
[arXiv:0704.4001 [hep-th]].

\bibitem{Demirtas:2019sip}
M.~Demirtas, M.~Kim, L.~Mcallister and J.~Moritz,
[arXiv:1912.10047 [hep-th]].

\bibitem{Linde:2020mdk}
A.~Linde,
[arXiv:2002.01500 [hep-th]].

\bibitem{Bellac:2011kqa}
M.~L.~Bellac,
doi:10.1017/CBO9780511721700

\bibitem{Chung:2011it}
D.~J.~Chung and A.~J.~Long,
Phys. Rev. D \textbf{84} (2011), 103513
doi:10.1103/PhysRevD.84.103513
[arXiv:1108.5193 [astro-ph.CO]].

\bibitem{Karwal:2016vyq}
T.~Karwal and M.~Kamionkowski,
Phys. Rev. D \textbf{94} (2016) no.10, 103523
doi:10.1103/PhysRevD.94.103523
[arXiv:1608.01309 [astro-ph.CO]].

\bibitem{Lyth:1995ka}
D.~H.~Lyth and E.~D.~Stewart,
Phys. Rev. D \textbf{53} (1996), 1784-1798
doi:10.1103/PhysRevD.53.1784
[arXiv:hep-ph/9510204 [hep-ph]].

\bibitem{Verde:2019ivm}
L.~Verde, T.~Treu and A.~Riess,
doi:10.1038/s41550-019-0902-0
[arXiv:1907.10625 [astro-ph.CO]].



\end{thebibliography}
\end{document}